\documentclass[useAMS,usenatbib,usegraphicx,aastex]{mn2e}

\def\aap{A\&A}%
\def\apjl{ApJ}%
\def\apj{ApJ}%
\def\aj{AJ}%
\def\apjs{ApJS}%
\def\mnras{MNRAS}%
\def\araa{ARA\&A}%
\def\nat{Nature}%
\def\pasp{PASP}%

\title[Optical observations of five GRBs]{Prompt, early, and
afterglow optical observations of five gamma-ray bursts (GRBs 100901A,
100902A, 100905A, 100906A, and 101020A)}
\author[E. S. Gorbovskoy et al]
{E. S. Gorbovskoy$^{1}$\thanks{E-mail:gorbovskoy@sai.msu.ru},
G. V. Lipunova$^{1}$\thanks{E-mail: galja@sai.msu.ru},
V. M. Lipunov$^{1}$, 
V. G. Kornilov$^{1}$, \newauthor
A. A. Belinski$^{1}$, 
N. I. Shatskiy$^{1}$, 
N. V. Tyurina$^{1}$,
D. A. Kuvshinov$^{1}$, \newauthor 
P. V. Balanutsa$^{1}$,
V. V. Chazov$^{1}$, 
A. Kuznetsov$^{1}$, 
D. S. Zimnukhov$^{1}$, \newauthor
M. V. Kornilov$^{1}$, 
A. V. Sankovich$^{1}$,
A. Krylov$^{1}$, 
K. I. Ivanov$^{2}$, 
O. Chvalaev$^{2}$,  \newauthor
V. A. Poleschuk$^{2}$, 
E. N. Konstantinov$^{2}$, 
O. A. Gress$^{2}$,
S. A. Yazev$^{2}$, 
N. M. Budnev$^{2}$, \newauthor
V. V. Krushinski$^{3}$, 
I. S. Zalozhnich$^{3}$, 
A. A. Popov$^{3}$,
A. G. Tlatov$^{4}$, \newauthor
A. V. Parhomenko$^{4}$, 
D. V. Dormidontov$^{4}$, 
V. Sennik$^{4}$,
V. V. Yurkov$^{5}$, \newauthor 
Yu. P. Sergienko$^{5}$, 
D. Varda$^{5}$, 
I. P. Kudelina$^{5}$, 
A. J. Castro--Tirado$^{6}$,\newauthor 
J. Gorosabel$^{6}$, 
R. S\'anchez--Ram\'irez$^{6}$, 
M. Jelinek$^{6}$, 
J. C. Tello$^{6}$\\
$^{1}$Sternberg Astronomical Institute, Moscow State University,  
Universitetskiy pr. 13, Moscow 119992, Russia\\
$^{2}$Irkutsk State University, ul. Karla Marxa 1, Irkutsk 664003, Russia\\
$^{3}$Kourovka Astronomical Observatory, Physical Department
of Ural State University, pr. Lenina 51, Ekaterinburg  620083, 
Russia\\
$^{4}$Kislovodsk Solar Station of the Pulkovo Observatory RAS, P.O.Box
45, ul. Gagarina 100, Kislovodsk 357700, Russia\\
$^{5}$Blagoveschensk Educational State University, ul. Lenina 104, 
Amur Region, Blagoveschensk 675000, Russia\\
$^{6}$Instituto de Astrof\'isica de Andaluc\'ia (IAA-CSIC), Glorieta de
la Astronom\'ia s/n, 18008 Granada, Spain
}
\begin{document}
\date{Accepted 2011 November 13. Received 2011 November 02; in original
form 2011 June 24}

\maketitle

\label{firstpage}

\begin{abstract}
We present results of the prompt, early, and afterglow optical
observations of five $\gamma$-ray bursts, GRBs 100901A, 100902A,
100905A, 100906A, and 101020A, made with the Mobile Astronomical System
of TElescope-Robots in Russia (MASTER-II net), the \hbox{1.5-m}
telescope of Sierra-Nevada Observatory, and the 2.56-m Nordic Optical
Telescope. For two sources, GRB\,100901A and GRB\,100906A, we detected
optical counterparts and obtained light curves starting before cessation
of $\gamma$-ray emission, at 113~s and 48~s after the trigger,
respectively. 
Observations of GRB\,100906A were conducted in two
polarizing filters. Observations of the other three bursts gave the
upper limits on the optical flux; their properties are briefly
discussed. More detailed analysis of GRB\,100901A and GRB\,100906A
supplemented by {\em Swift} data provides the following results and
indicates a different origin of the prompt optical radiation in the two
bursts. The light curves patterns and spectral distributions suggest a
common production site of the prompt optical and high-energy emission in
GRB\,100901A. Results of spectral fits for GRB\,100901A in the range
from the optical to X-rays favor power-law energy distributions and a 
consistent value of the optical extinction in the host galaxy. 
GRB\,100906A produced a smoothly peaking optical light curve suggesting 
that the prompt optical radiation in this GRB originated in a front
shock. This is supported by a spectral analysis. We have found that the
Amati and Ghirlanda relations are satisfied for GRB\,100906A. An upper
limit on the value of the optical extinction on the host of GRB\,100906A
is obtained. 
\end{abstract}

\section{Introduction}
\begin{table*}
 \centering
  \caption{Photometry of GRB\,100901A by MASTER: $R$, unfiltered $W$,
$I$, and $V$.
Times $t-T0$ are the middle times
of exposures, which are listed in the 2nd, 5th, 8th, and 11th column.
Full  table can be found online in the electronic issue.
}
\label{tab:table_lc100901}
  \begin{tabular}{@{}llllllllllll@{}}
\hline
$t-T0$ & Exp. & $R$  &
$t-T0$ & Exp. & $W$  &
$t-T0$ & Exp. & $I$  &
$t-T0$ & Exp. & $V$  \\
 (h)&  (s) &  (mag) &
 (h)&  (s) &  (mag) &
 (h)&  (s) &  (mag) &
 (h)&  (s) &  (mag) \\
\hline
4.1381	& 180 &	$	18.08	\pm	0.45	$	&	0.0315	& 20&	$	18.93	\pm	2.00	$	&	5.4149	& 180 &      $       17.00   \pm     0.09    $       &       0.3811  & 180 &      $       17.96   \pm     0.19    $       \\
4.3939	& 180 &	$	17.81	\pm	0.17	$	&	0.0426	& 30&	$	18.23	\pm	0.80	$	&	5.4719	& 180 &      $       16.96   \pm     0.09    $       &       0.5449  & 180 &      $       17.96   \pm     0.19    $       \\
4.4545	& 180 &	$	17.56	\pm	0.11	$	&	0.0591	& 40&	$	18.42	\pm	0.80	$	&	5.5288	& 180 &      $       17.07   \pm     0.10    $       &       0.6021  & 180 &      $       17.76   \pm     0.17    $       \\
4.5129	& 180 &	$	17.61	\pm	0.10	$	&	0.0950	& 60&	$	18.53	\pm	0.70	$	&	5.5857	& 180 &      $       16.87   \pm     0.08    $       &       5.5865  & 180 &      $       17.59   \pm     0.11    $       \\
4.5705	& 180 &	$	17.45	\pm	0.08	$	&	0.1185	& 80&	$	17.51	\pm	0.25	$	&	5.6426	& 180 &      $       17.04   \pm     0.09    $       &       5.6451  & 180 &      $       17.70   \pm     0.12    $       \\
4.6278	& 180 &	$	17.77	\pm	0.10	$	&	0.1475	& 100 &	$	18.47	\pm	0.50	$	&	5.6999	& 180 &      $       17.04   \pm     0.09    $       &       7.1125  & 180 &      $       17.92   \pm     0.18    $       \\
4.6857	& 180 &	$	17.33	\pm	0.07	$	&	0.1820	& 120&	$	18.85	\pm	0.70	$	&	5.7568	& 180 &      $       17.05   \pm     0.09    $       &       7.1718  & 180 &      $       17.79   \pm     0.15    $       \\
4.7428  & 180 & $       17.56   \pm     0.09    $       &       0.2233  &  150& $       18.62   \pm     0.60    $       &       5.8138  & 180 &      $       16.83   \pm     0.07    $       &       	     &     	& $       $       \\
4.8000  & 180 & $       17.45   \pm     0.07    $       &       0.2732  &  180 &$       17.94   \pm     0.30    $       &       5.8706  & 180 &      $       16.81   \pm     0.07    $       &       		&       & $       $       \\
4.8571  & 180 & $       17.39   \pm     0.07    $       &       0.3272  & 180 & $       17.99   \pm     0.30    $       &       5.9277  & 180 &      $       16.92   \pm     0.07    $       &       		&       & $       $       \\
\hline    
\end{tabular}
\end{table*}

Since 1997, when the optical radiation of $\gamma$-ray bursts (GRBs) was
first detected, it is known that they are the most energetic events in the
Universe~\citep{Kulkarni_etal1998}. 
Optical emission observed hours after a GRB is attributed to the
so-called afterglow, which is the result of a shock propagating outward
in the surrounding media~\citep{meszaros-rees1997}.
Characteristics of such emission are defined mainly by conditions in the
interstellar media and the amount of the released energy but depend
weakly on details of the central burst. 

 The physics of GRBs and emission mechanisms are not completely
understood. A further progress calls for more observational data and
model analysis. An acknowledged model is that a GRB is a manifestation
of a formation of a rotating black hole (or another compact relativistic
object) in a course of the gravitational collapse. An engine works that
converts energy of the collapse into emissions of different types, among
which a high-energy $\gamma$-emission is produced for up to several tens
of seconds. The afterglow emission is detected long afterwards, while
the engine is still working or not, which is uncertain. In order to
understand better the details of the process, it is necessary to observe
the main event itself, at different wavebands, while the engine is at
its most active stage.

However, to observe prompt optical emission is a challenge as a GRB
usually lasts no more than several tens of seconds. Beginning with the
prompt optical observations of GRBs in 1998 of~\citet{akerlof_etal2000},
successful prompt optical detections remain rare. Evidently, two
approaches can be conceived: to observe extensive sky fields waiting for
a GRB to occur or to use special robotic telescopes ready to point
anywhere by an alert from an orbital $\gamma$-ray observatory--`alert
observations' technique. The Mobile Astronomical System of
TElescope-Robots in Russia (MASTER-II\footnote{MASTER web site:
http://observ.pereplet.ru}) use both
techniques~\citep{lipunov_etal2010}. The present paper is dedicated to
the alert MASTER observations of five GRBs in Siberia, Ural, and North
Caucasus.

Alert ground-based optical observations of GRBs are a new global
physical experiment available since the last decade.
It is made possible thanks to the
implementation of the global Internet network, powerful personal
computers, and fast optical CCD-receivers. 
The challenge is to
accomplish quickly the four following steps as early as possible:
\begin{enumerate}
\item  a GRB is detected by a $\gamma$-ray telescope on board
of a spacecraft ({\em Swift} \citep{Swift2004}, {\em
Fermi}~\citep{Fermi2009}, {\em INTEGRAL}~\citep{Integral2003}, etc.) 
\item after on-board processing is finished, location of a GRB
is sent to the Gamma-ray bursts Coordinates Network (GCN) at NASA. The
first two steps take from 10 to 40 seconds.
\item the burst position is then disseminated to ground-based robotic
telescopes through the Internet network -- in about 0.5 s.
\item robotic telescopes are scheduled and pointed to the received 
positions, which takes from 7 to 40 s for moderate-size instruments 
(less than 0.5 m) and from several minutes to hours for 2m- and larger
instruments. Whereupon, imaging is made in optical and IR bands.
\end{enumerate}

The first Russian robotic telescope MASTER (Mobile Astronomical System
of Telescope-Robots) came into operation in 2002 near Moscow thanks to
the private funding of Moscow association 
`Optics'\footnote{http://www.ochkarik.ru/master/}. Construction of the
all-Russia network MASTER began in 2008 \citep{lipunov_etal2010}.
Presently, telescopes of the MASTER-net are located in observatories of
Moscow State University (in Kislovodsk), Ural State University (in
Kourovka), and Irkutsk State University (in Tunka near Baikal Lake) and
the Blagoveschensk Pedagogical University (in Blagoveschensk region).
These observatories span six time zones.
Description of the MASTER II telescopes can be
found in \citet{kornilov_etal2011}.

In September and October 2010, the MASTER telescopes made five target
observations of GRBs that triggered the {\em Swift} observatory: GRBs
100901A, 100902A, 100905A, 100906A, and 101020A. In the present paper,
we report these observations and make an analysis for two long GRBs,
whose counterparts were successfully detected. For one of them,
GRB\,100906A, we also use the data obtained with the \hbox{1.5-m}
telescope at Sierra-Nevada Observatory (OSN) and \hbox{2.56-m} Nordic
Optical Telescope (NOT). 
 
We begin in Section~\ref{section:obs} with a description of the
observations. Reduction and analysis of the data obtained by MASTER are
described in Section~\ref{section:methods}. Preparation of {\em Swift}
light curves and spectra from the data available online is also
described there. In Section~\ref{section:results} we turn to the results
for GRBs\,100901A and 100906A: the spectra before $T_{90}$, the
late-time spectra and the estimates of the optical extinction in the
host galaxies, the search of the jet break time on the GRB\,100906A
light curve after 10\,000~s. In Section~\ref{section:discussion} we
consider the arguments for different sites of the prompt optical
emission in GRBs\,100901A and 100906A. We also discuss their spectral
evolution, and the spectral characteristics of other GRBs, for which
optical counterparts were not detected. The applicability of the Amati
and Ghirlanda relations to GRBs\,100901A and 100906A is checked. We also
mention the model of a two-step collapse for the long engine activity
that can cause a bump on the X-ray and optical light curve of
GRB\,100901A. We summarize our results in Section~\ref{section:summary}.

Throughout this work, we adopt a standard CDM cosmology with $H_0 = 70
\,\mathrm{km}\, s^{-1}\, \mathrm{Mpc}^{-1}$, $\Omega_\mathrm{m} = 0.27$, and
$\Omega_\mathrm{\Lambda}= 0.73$.
We define the flux density power-law temporal and spectral decay indices
as $F_\nu \propto t^{-\alpha} \nu^{-\beta}$, and the photon index 
equals $\beta+1$.
All errors in the paper are 1$\sigma$ uncertainties unless otherwise
noted. All {\em Swift} data used to reproduce light curves and spectra
were taken from the {\em Swift}
Catalog\footnote{http://heasarc.nasa.gov/docs/swift/archive/grb\_table/}, 
the UK {\em Swift} data
archive\footnote{http://www.swift.ac.uk/swift\_portal/} and the Burst
Analyser website.\footnote{http://www.swift.ac.uk/burst\_analyser/}

\section{Observations}
\label{section:obs}

Presently the MASTER II net consists of four identical, fast telescopes
located at different sites in Russia
\citep{lipunov_etal2010,kornilov_etal2011}. Each telescope is a
twin-tube instrument with the aperture of 0.4~m and the focal ratio
f/2.5. Each telescope is equipped with the CCD camera $4\times4$K
AltaU-16M providing a field of view of 2x2 square degrees. A photometric
unit provides a set of four filters, which can be placed in turn at the
optical axis before the CCD camera. Depending on the long-term
scientific task, the photometric units of the twin telescopes are
supplied with three filters (any of the following: Johnson $B$, $V$,
$R$, $I$, and white glass). The fourth position is occupied by
polarization filters, oriented orthogonally in the two tubes.
Polarization filters are positioned differently in the celestial
coordinate system at the telescopes of the MASTER net.

Each observatory operates automatically. Waiting for an alert signal,
MASTER II monitors the sky searching for possible optical transients.
Swapping from the monitoring to alert observations of GRBs is
accomplished by switching the filters, focusing, and pointing the
telescope to the event coordinates received from the GCN. In 20 seconds
after an alert, the MASTER II observatory is ready to observe a GRB. 
The pointings at the positions of the GRBs considered below took from 22
to 47~s. 

\subsection{Prompt optical observations of GRB\,100901A}
\label{s:obs_100901a}

GRB\,100901A triggered the {\em Swift} Burst Alert Telescope ({\em
Swift}/BAT) at 13:34:10 UT on September 1, 2010. Time $T_{90}$, during
which $90$ per cent of $\gamma$-fluence is detected, was $439 \pm
33$~s~\citep{gcn11159, gcn11169b}.

Two MASTER robotic telescopes were pointed to the {\em Swift} BAT
position of GRB\,100901A: at Tunka at 103~s after the BAT trigger time
$T0$ and near Blagoveschensk, 101~s after $T0$ \citep{gcn11161}. We have
got more than \hbox{$\sim 5$~min} of optical observations simultaneous
to the prompt $\gamma$-emission of the GRB. The source was too low above
the horizon at Tunka, and it was possible to observe it with only one of
the optical tubes. The second tube was shadowed by the dome, which made
impossible measuring polarized light. The eastern telescope, near
Blagoveschensk, was not operational because of the bad weather
conditions. An optical flare, with maximum at 17.0$^\mathrm{m}$ in the
MASTER unfiltered band, was clearly detected at $426\pm 40$~s after the
trigger time~\citep{gcn11163}, synchronously with the X-ray
flare~\citep{gcn11171} and the $\gamma$-flare~\citep{gcn11169b}. 

\begin{figure}
\includegraphics[width=84mm]{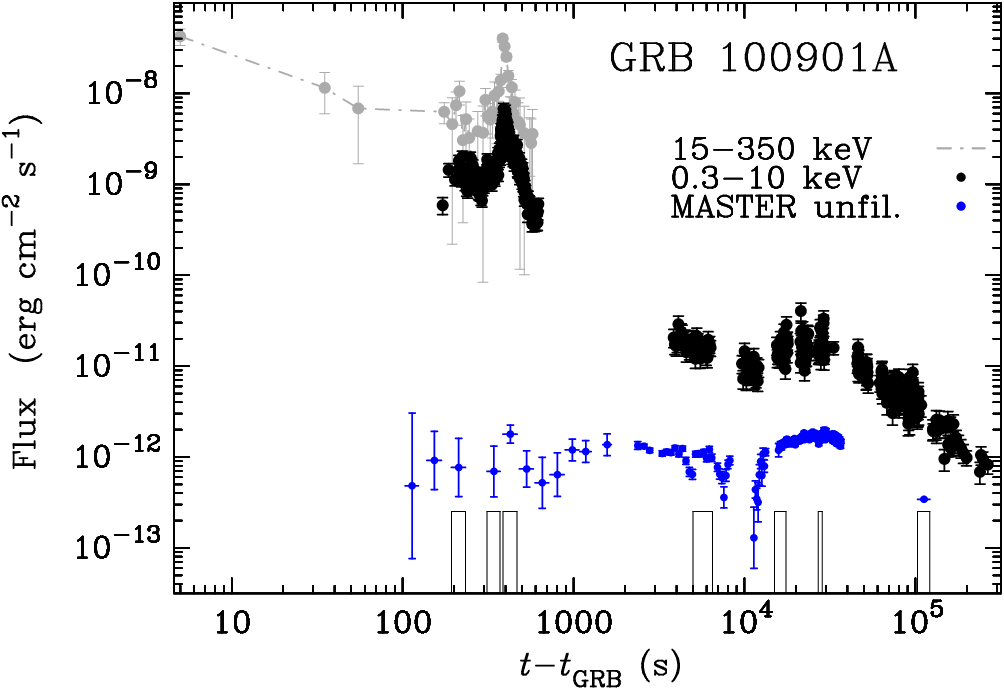}
\caption{GRB\,100901A optical light curve obtained by MASTER (blue
points with vertical bars representing errors and horizontal bars
representing exposure times; data are corrected for the Galactic
extinction) along with the {\em Swift}/BAT 15-350~keV flux with 10~s
binning~(grey dots connected by dot-dashed lines) and {\em Swift}/XRT
0.3--10~keV unabsorbed flux (black dots; \citealt{gcn11171}).  Thin-line
rectangles show the time intervals selected for spectral analysis. Color
figures can be viewed in the electronic version.
\label{fig:lc100901a_all}}
\end{figure}

\begin{figure}
\includegraphics[width=84mm]{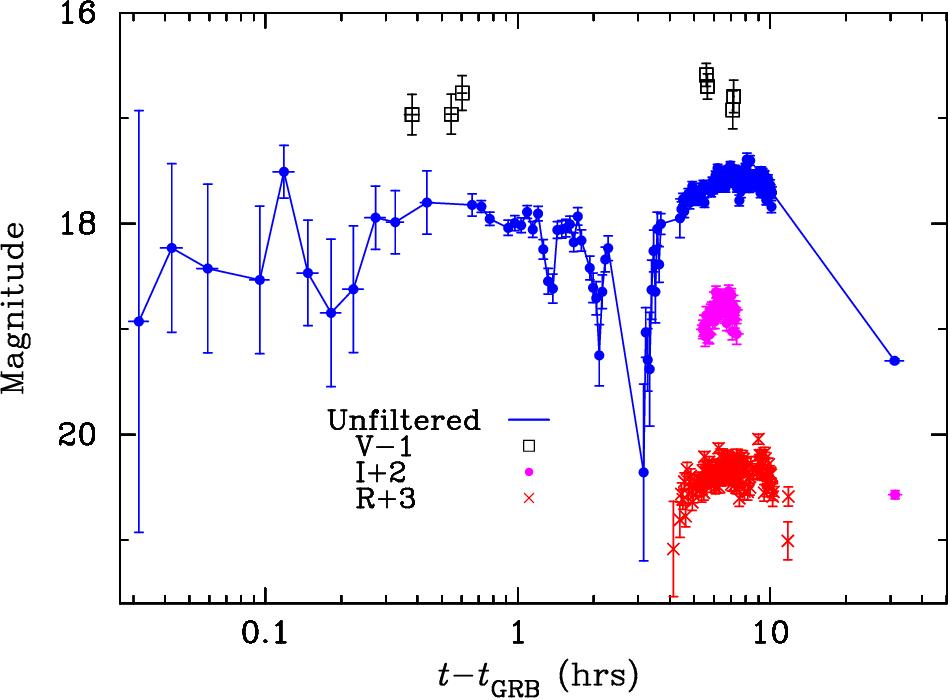}
\caption{MASTER optical light curves of GRB\,100901A in different
filters: $R$ (3 mag added; red crosses), $I$ (2 mag added; magenta
dots), $V$ (1 mag subtracted; black squares), 
and unfiltered (blue dots connected with solid line).
Magnitudes are not corrected for the Galactic extinction.
Horizontal bars show exposures.
\label{fig:lc100901a_opt}}
\end{figure}

The MASTER telescope at Kourovka was pointed to GRB\,100901A $\sim 5$
hours later. Overall, we have been continuously observing the GRB during
14 hours at three sites, Tunka, Kourovka, and Kislovodsk, in different
optical bands: unfiltered, $R$, $I$, and $V$. The resulting data are
summarized in Table~\ref{tab:table_lc100901} and presented in
Figs.~\ref{fig:lc100901a_all} and \ref{fig:lc100901a_opt}. In
Fig.~\ref{fig:lc100901a_all}, {\em Swift}/XRT 0.3-10~keV and {\em
Swift}/BAT 15-350~keV light curves for GRB\,100901A are shown as well
(for description see Section~\ref{section:bat_xrt}). 
 
OSN observed the GRB about 11 hours after the trigger~\citep{gcn11180}.
The redshift of the host galaxy z=1.408 was obtained at the 8-m
Gemini-North telescope~\citep{gcn11164}.

\subsection{Prompt optical observations of GRB\,100902A}

On September the 2nd, the Kourovka MASTER telescope was pointed to the
BAT position of GRB\,100902A \citep{gcn11181c} 104~s after the GRB
trigger time $T0$ 19:31:54 UT. This GRB flaring activity lasted for
about 200~s after the trigger time, and MASTER was first to witness its
active stage in the optical waveband. The imaging was attempted at two
polarization angles with 20~s exposure times. However, no optical
transient was found at the XRT position with the upper limit of
$17$~mag~\citep{gcn11182}. The final results of our
photometry~\citep{gcn11185}, which are the upper limits on the optical
flux at two polarization angles, are summarized in
Table~\ref{tab:100902_limits}. The GRB position was observed by the OSN
at $\sim 7$~h after the trigger with an upper limit on the magnitude in
$I$-band: $\sim 21.8$~mag~\citep{gcn11196}.

\begin{table}
 \centering
  \caption{MASTER upper limits on the optical flux obtained with
the polarizing filter for GRB\,100902A. The bottom row is coadded from
all above.}
\label{tab:100902_limits}
  \begin{tabular}{@{}cccccc@{}}
  \hline
$t_\mathrm{start}$ & 
$t_\mathrm{end}$    &   
Exposure time & 
$t_\mathrm{start}-T0$ &  
mag \\
(UT) & 
 (UT)   &   
(s) & 
 (s) &  
 \\
\hline
19:33:38 & 19:33:58&      20      &    104.0 & 17.0   \\
19:34:12 & 19:34:42&      30      &    132.0 & 17.3  \\
19:34:57 & 19:35:57&      40      &    177.0 & 17.3  \\
19:35:51 & 19:36:41&      50      &    231.0 & 17.4  \\
19:33:38 & 19:36:41&     140      &    104.0 & 18.0  \\
\hline
\end{tabular}
\end{table}

\subsection{Early optical observations of GRB\,100905A}

On September the 5th, a GRB was detected by {\em Swift}/BAT at 15:08:15
UT~\citep{gcn11214}. The Tunka MASTER telescope was pointed to its
position 55~s after $T0$. The imaging was carried out at two
polarization angles~\citep{gcn11216}.
\begin{table}
\centering
  \caption{MASTER upper limits on the optical flux obtained with
polarizing filter for GRB\,100905A.
}
\label{tab:100905_limits}
  \begin{tabular}{@{}ccccc@{}}
  \hline
$t_\mathrm{start}$ & $t_\mathrm{end}$   &   
Exposure time&  
$t_\mathrm{start}-T0$&
mag \\
(UT)   &   
(UT) &  
 (s) &(s) &
\\
\hline
15:09:10& 15:09:20& 10&  55.0& 16.5\\
15:09:39& 19:09:59& 20&  84.0& 17.1\\
15:10:45& 15:11:15& 30& 150.0& 17.3\\
15:11:35& 15:12:15& 40& 200.0& 17.5\\
\hline
\end{tabular}
\end{table}
No optical transient is found. Upper limits are given in
Table~\ref{tab:100905_limits}. 
The further optical observations were carried out by
UVOT~\citep{gcn11237} and the 1.5-m telescope of the Observatorio de
Sierra Nevada~\citep{gcn11220} and yielded no optical counterpart
either. \citet{gcn11218} obtained that $T_{90}=3.4\pm0.5$~s implying
that our observations can be qualified as `early', not prompt. This
$\gamma$-ray burst might belong to the class of intermediate duration
GRBs, separated from the class of long GRBs on statistical grounds by
\citet{horvath1998,horvath2002,Mukherjee_etal1998}. While underlying
physical properties of the subclasses of long GRBs appear to be similar
(type of progenitor, interstellar environment), there is an evidence for
duration and luminosity differences caused by some slight diversity of
the bursts~\citep{postigo_etal2011}.

\subsection{Prompt and afterglow optical observations of GRB\,100906A}

At 13:49:27 UT, September the 6th, GRB\,100906A was
detected~\citep{gcn11227}. The BAT/{\em
Swift} detected  several bright peaks with 
 $T_{90}=114.4\pm 1.6$~s \citep{gcn11233}.
GRB\,100906A has been also observed by the
Konus--{\em Wind} starting from 13:49:30 UT~\citep{gcn11251}. 

Two MASTER telescopes, at Tunka and Blagoveschensk, were pointed to the
GRB position, respectively, 38~s after the BAT trigger time $T0$ and
58~s after $T0$. As a pilot instrument was mounted at the time at the
Blagoveschensk site\footnote{Since December 2010, the Blagoveschensk
telescope is in a full operating mode.}, we present only the data
obtained by the telescope at Tunka. In Kourovka and Kislovodsk, the
weather conditions were not suitable for observations.

\begin{figure}
\includegraphics[width=84mm]{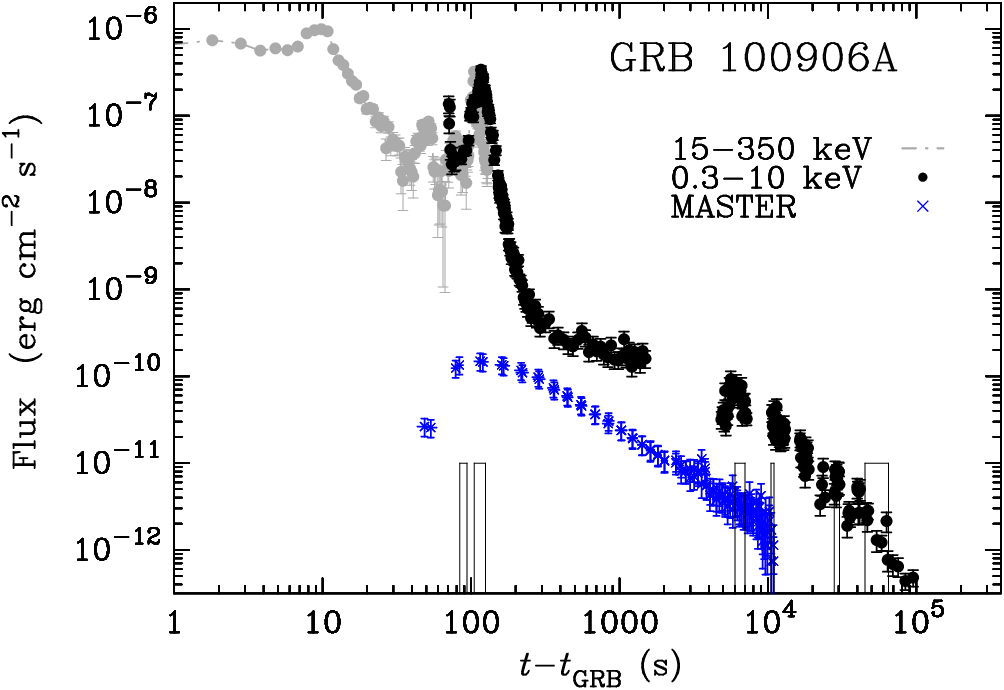}
\caption{GRB\,100906A optical light curve obtained by MASTER with one
polarization filter (blue dots with 20 per cent error bars resulted due to 
uncertainty of magnitude-flux conversion, see 
\S~\ref{section:flux_calibration}; data are corrected for the Galactic
extinction) along with the {\em Swift}/BAT
15-350~keV flux with 1 s binning (grey dots connected with dot-dashed
line; data with negative lower limits are not shown) and the  {\em
Swift}/XRT 0.3-10~keV unabsorbed-flux light curve (black dots; \citealt{gcn11244}). 
Thin-line rectangles show the time intervals selected for spectral
analysis.
\label{fig:lc100906a_all}}
\end{figure}

\begin{figure}
\includegraphics[width=84mm]{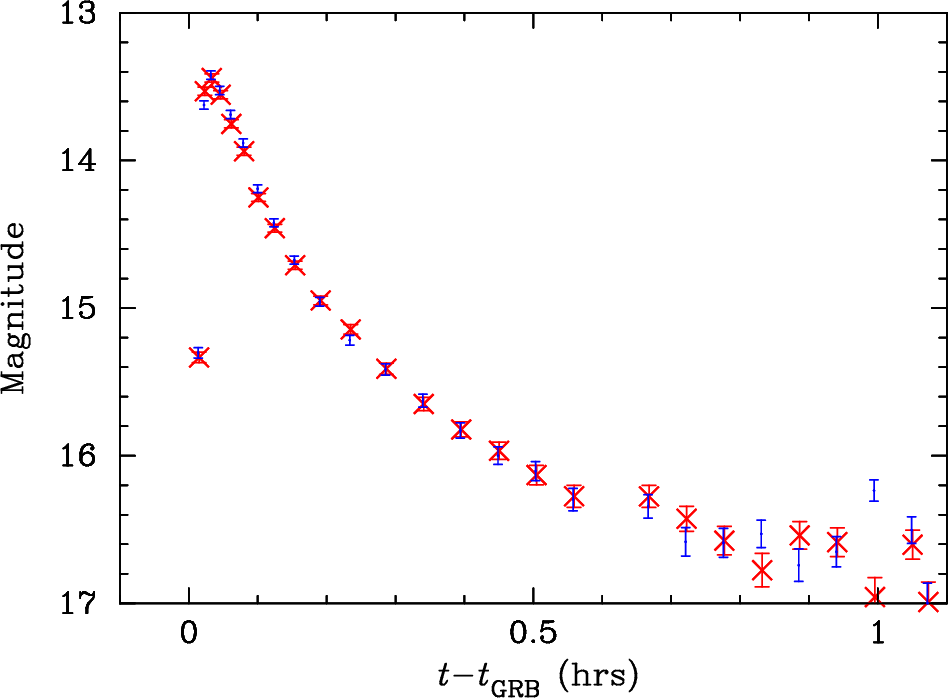}
\caption{MASTER light curve of GRB\,100901A in two polarizations,
unfiltered band (blue dots and red crosses). Data are not corrected for
the Galactic extinction. 
\label{fig:lc100906a_opt}}
\end{figure}

A bright optical transient was localized by MASTER at the {\em
Swift}/UVOT position~\citep{gcn11227}, 13 mag at maximum. During the
first hour after the trigger time, 24 images in unfiltered light at two
polarization angles were produced by the Tunka telescope.
Fig.~\ref{fig:lc100906a_all} shows the early observations of
GRB\,100906A in optical, 0.3-10~keV, and 15-150~keV energy range (also
Fig.~\ref{fig:lc100906a_opt} and Table~\ref{tab:polarization100906}). A
movie of the optical burst of GRB\,100906A as seen at Tunka is available
on the web\footnote{http://master.sai.msu.ru/static/GRB/grb100906.avi}.

We calculate the relative difference between the signals coming from the
two polarization filters, for the time interval $100-10^4$~s after T0.
It is less than the relative measuring accuracy of filters 2 per cent,
estimated through the observations of standard stars. Unfortunately,
technical limitations do not allow us to draw a decisive conclusion
about the polarization degree of GRB\,100906A~(see
\S~\ref{section:opt_sites} and the Appendix). 

\begin{table}
 \centering \caption{Photometry data for GRB\,100906A by MASTER in the
unfiltered band with two orthogonal polarizing filters.  
Full table can be found online in the electronic issue. 
}
\label{tab:polarization100906}
  \begin{tabular}{@{}cccccc@{}}
\hline
\multicolumn{3}{c}{$P_{\updownarrow}$}&\multicolumn{3}{c}{$P_{\leftrightarrow}$}\\
  \hline
$t-T0$&Exp.&Unfiltered&$t-T0$
&Exp.&Unfiltered\\
 (h)  &       (s)     & (mag)    & (h)  &  (s)   &(mag) \\
\hline
0.013495 &	 10  & $ 	 15.30  \pm	 0.04  $ &	\dots  & \dots  & $ \dots   $ \\	
0.021896 &	 10  & $ 	 13.62  \pm	 0.03  $ &	0.023187  &	 10  & $ 	 13.53  \pm	 0.03  $ \\	
0.032043 &	 20  & $ 	 13.42  \pm	 0.03  $ &	0.033377  &	 20  & $ 	 13.44  \pm	 0.03  $ \\	
0.044763 &	 30  & $ 	 13.53  \pm	 0.03  $ &	0.046099  &	 30  & $ 	 13.55  \pm	 0.03  $ \\	
0.060409 &	 40  & $ 	 13.69  \pm	 0.03  $ &	0.061662  &	 40  & $ 	 13.75  \pm	 0.03  $ \\	
0.078794 &	 50  & $ 	 13.88  \pm	 0.03  $ &	0.080028  &	 50  & $ 	 13.94  \pm	 0.03  $ \\	
0.099815 &	 60  & $ 	 14.19  \pm	 0.03  $ &	0.101093  &	 60  & $ 	 14.25  \pm	 0.03  $ \\	
0.123506 &	 70  & $ 	 14.42  \pm	 0.03  $ &	0.124784  &	 70  & $ 	 14.46  \pm	 0.03  $ \\	
0.152931 &	 90  & $ 	 14.68  \pm	 0.03  $ &	0.154302  &	 90  & $ 	 14.71  \pm	 0.03  $ \\	
0.189999 &	 110  & $ 	 14.96  \pm	 0.03  $ &	0.191257  &	 110  & $ 	 14.95  \pm	 0.03  $ \\	
\hline    
\end{tabular}
\end{table}

\begin{figure}
\includegraphics[width=84mm]{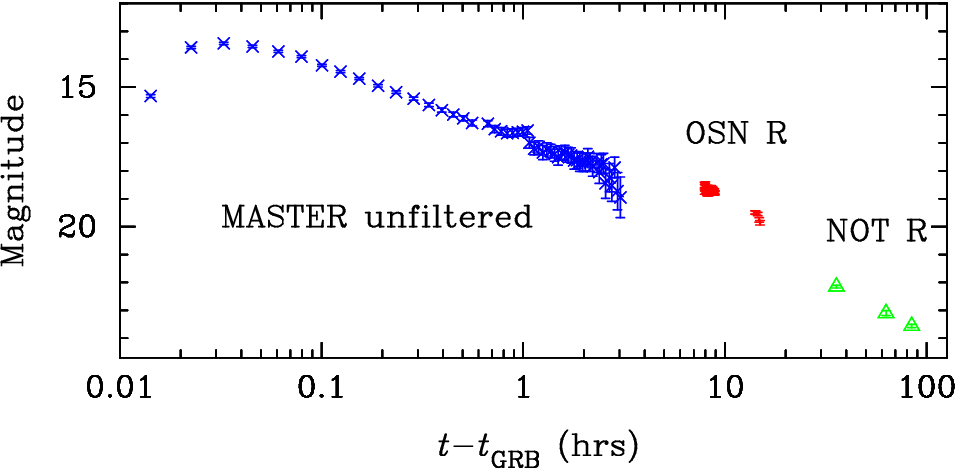} \caption{ OSN
and NOT telescopes $R$-band light curve of GRB\,100906A (red bars and
green triangles with small error bars, respectively). MASTER unfiltered
band, one polarization light curve is plotted by blue crosses before
5~h. All data are not corrected for the Galactic extinction. 
\label{fig:lc100906a_opt_OSN}}
\end{figure}

The 1.5m OSN telescope at Sierra Nevada observed the GRB few hours
after $T0$ (Fig.~\ref{fig:lc100906a_opt_OSN} and 
Table~\ref{tab:OSN100906}). Calibration for $BVIR$
bands is done using 49 USNO stars, and for $U$-band using a 
Landolt star observation. Aperture photometry is carried out using {\tt
phot} as implemented in IRAF with a radius equal to the seeing. The
error of the afterglow magnitude was obtained adding quadratically the
statistical error given by {\tt phot} and the zero point error 
determined with the 49 field reference stars.
In addition, Table~\ref{tab:NOT100906} shows the late photometry
 obtained by the 2.56-m Nordic Optical Telescope (NOT).

The redshift of the host galaxy $z=1.727$ was obtained with the
Gemini-North telescope~\citep{gcn11230}.

\subsection{Limits on the prompt and afterglow optical flux of
GRB\,101020A}

\begin{table}
 \centering \caption{$UBVRI$-photometry for GRB\,100906A obtained by the
OSN telescope. Reported times are middle times of exposures. Dots
substitute 50 observations that can be found online in the electronic
version.}
\label{tab:OSN100906}
  \begin{tabular}{@{}cccc@{}}
\hline
Band&$t-T0$&Exposure time&Magnitude \\
    & (h)  &  (s)        &\\
\hline
R	& 7.920835 &	 36  & $ 	 18.53  \pm	 0.12  $  	\\
R 	& 7.935742 &	 36  & $ 	 18.72  \pm	 0.12  $  	 \\
$\dots$ &   $\dots$  &   $\dots$ &	$\dots$				  \\
 R 	& 8.895094 &       60  & $         18.79  \pm      0.06  $ \\
 R 	& 14.091390 &      180  & $        19.49  \pm      0.06  $ \\
 R 	& 14.217964 &      180  & $        19.50  \pm      0.06  $ \\
 R 	& 14.540464 &      180  & $        19.55  \pm      0.06  $ \\
 R 	& 14.784353 &      180  & $        19.76  \pm      0.08  $ \\
 R 	& 14.928057 &      180  & $        19.86  \pm      0.08  $ \\ 
 R 	&  38.528342 &    2$\times$300  & $>21.5 $  \\      
 V      &     14.141108 &     180  & $        19.91  \pm     0.20  $ \\    
 V      &     38.402228 &     300  & $        >21.0   $ \\
 B      &     14.193888 &     300  & $        20.65  \pm     0.22  $ \\    
 B      &     38.318343 &     5$\times$300  & $       >21.0           $ \\
 I      &     14.278609 &     2$\times$180  & $       19.46  \pm     0.12  $ \\
 U      &     14.571669 &     2$\times$900  & $       >17.5           $ \\
 U      &     38.234160 &     300  & $        >16.5           $        \\
\hline    
\end{tabular}
\end{table}

\begin{table}
 \centering
 \caption{R-magnitudes of GRB\,100906A obtained by NOT.}
\label{tab:NOT100906}
  \begin{tabular}{@{}ccccccc@{}}
\hline
$t_\mathrm{start}-T0$&
$t_\mathrm{end}-T0$& 
Exp. time&$R$&\\
(h)&
(h) & 
(s) &(mag)&\\
\hline
35.031028 & 36.338862 &  $3\times 600$ &  $22.15\pm 0.05 $  \\
62.828250 & 63.078251 &  $900$         &  $23.10 \pm 0.09$  \\
84.046943 & 84.763888 &  $2\times 900$ &  $23.56 \pm 0.06$  \\
\hline    
\end{tabular}
\end{table}
On October the 10th, at 23:40:41 UT, the {\em Swift}/BAT triggered and
located GRB\,101020A~\citep{gcn11357}. \citet{gcn11358a} found
$T_{90}=175\pm28$~s. MASTER telescope at Kourovka was pointed to the
GRB\,101020A position 106~s after $T0$ at a large zenith distance of 80
degrees~\citep{gcn11359,gcn11361}. It was observed with two optical
tubes. No optical transient is detected with the upper limits in
$V$-band reported in Table~\ref{tab:101020_limits}, where magnitudes are
obtained from coadded images.
\begin{table}
 \centering \caption{Upper limits on the optical flux in $V$-band for
GRB\,101020A obtained for coadded images.}
\label{tab:101020_limits}
  \begin{tabular}{@{}ccccc@{}}
  \hline
$t_\mathrm{start}-T0$&   $t_\mathrm{middle}$ & Exp. time
&   $V$ & Coadded\\
(s)&   (s)& (s)&   (mag) & number\\
\hline
 106 &  116 &   20+20& 14.9   & 1+1   \\
 472 & 2836 &    4300&  18.5   & 25 \\
\hline
\end{tabular}
\end{table}

\section{Data reduction and analysis for GRB\,100901A and
GRB\,100906A}\label{section:methods}

\subsection{MASTER astrometric and photometric calibration }

Each MASTER telescope is equipped with a set of Johnson/Bessell 
filters and two linear polarizing laminated films.
For every image, astrometric calibration using {\tt imcoords} of IRAF
v2.14 is performed. Aperture photometry is done with the {\tt phot} of
IRAF package \citep{tody1993}. Standard stars from $13.5^\mathrm{m}$ to
$17^\mathrm{m}$ are selected from the Sloan Digital Sky Survey Data
Release 7 (SDSS-DR~7) \citep{abazajian_etal2009}. Influence of
atmospheric transparency variations on instrumental stellar magnitudes
is eliminated following an algorithm described in
\citet{everett-howell2001}. Stars with variations exceeding threefold
the errors calculated by IRAF from signal-to-noise information are
expelled from the list, and the remaining standard stars are related to
SDSS-DR7 data as suggested by
Lupton~(2005)\footnote{http://www.sdss.org/dr7/algorithms/sdssUBVRITransform.html}:
\begin{eqnarray}
B&=&g+0.1884\,(u-g)+0.1313\, ,\\
V&=&g-0.5784\,(g-r)-0.0038\, , \nonumber \\ 
R&=&r-0.2936\,(r-i)-0.1439\, ,\nonumber \\  
I&=&i-0.3780\,(i-z)-0.3974\, , \nonumber 
\label{eq:sdss_positions}
\end{eqnarray}
where $BVRI$ are the Johnson system and $i,~r,~g, \mathrm{and}~z$ 
are the Sloan Survey photometric system  (all in the Vega system). Resulting lists of standard
stars are given in Table~\ref{tab:st_stars_100901} for GRB\,100901A and
in Table~\ref{tab:st_stars_100906} for GRB\,100906A. Associated
variances$^6$ between observed and catalogue values after applying the
above relations in different filters are $0.005-0.008^\mathrm{m}$. In
our observations, standard deviations of light curves of the standard
stars are greater, $0.02-0.1$~mag depending on the filter.
For images obtained in the unfiltered band, we use a zero-point
correction to the unfiltered magnitudes obtained at the Kislovodsk MASTER
telescope. Flux calibration of the unfiltered magnitudes at Kislovodsk
is described in Section~\ref{section:flux_calibration}.

 The GRB magnitude errors are defined by the standard deviation of
the brightness variations of the standard stars of similar magnitude in
the case of high signal-to-noise ratio. At the low signal-to-noise
ratio, we lack standard stars of such magnitudes. To estimate
observational errors, we take the theoretical limit for errors following
the basic CCD equation. This limit is much greater (up to 2 mag at mag
$> 18$) than expected amplitude variations of standard stars at such
signal-to-noise level.

 For each of the standard stars, we evaluate the difference between its
average magnitudes in both orthogonal polarizations. Assuming that the
light of standard stars should not be polarized, the standard stars with
outlier polarization are expelled from the list. Afterwards, corrections
are applied to match average magnitudes of the standard stars in two
polarizations. The relative accuracy of observations with two filters is
approximately 2 per cent, basing on observations of standard stars.

\subsection{MASTER flux calibration\label{section:flux_calibration}}
\begin{figure*}
\includegraphics[width=140mm]{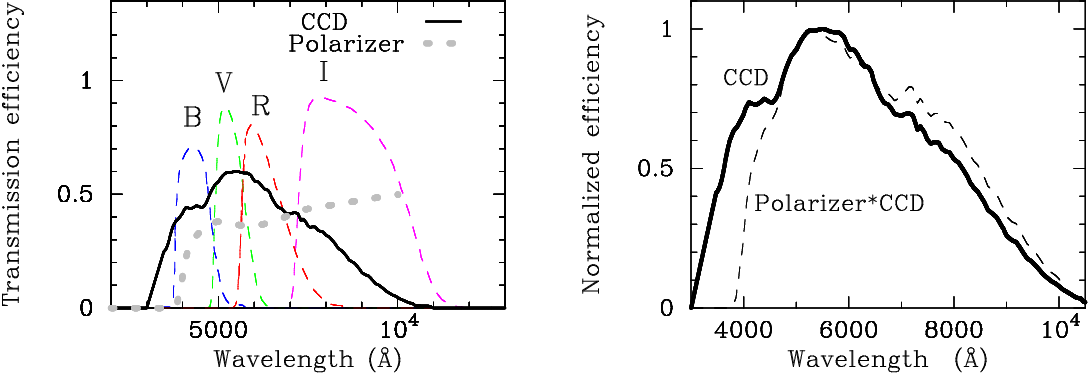} 
\caption{ 
In the left panel the  MASTER filter transmission efficiencies
are plotted along with those of the CCD  and  polarizer.
In the right panel, functions $W(\lambda)$ and $P(\lambda)$ are shown,
which are
the bandpasses of the CCD  and of the CCD plus polarizer, respectively.
\label{fig:calibration} }
\end{figure*}

To convert magnitudes $BVRI$ into absolute values, we use the
zero-magnitude flux densities of the Landolt photometry~\citep[see,
e.g.,][]{Pickles-Depagne2010}. To convert CCD photometry into absolute
fluxes without color filter interposed, that is, in a wide waveband (see
Fig.~\ref{fig:calibration}), we use the Pogson formula relating stellar
magnitudes and CCD band-integrated flux with a zero-magnitude flux
calculated using the Vega spectrum as given by \citet{Hayes1985}.

The standard stars are designated with the $W$ and $P$ magnitudes:
obtained from observations in the wide waveband and in the wide waveband
plus polarizer. Both are approximated by the formula $0.8\,R +0.2\,B$.
This formula is arrived at by compilation of color diagrams $W-X$ versus
$B-R$, where $X$ is a combination of stellar magnitudes, $X=a\,R+b\,B$.
In a general case, points group around a straight oblique line. The
zero-slope of the relation is achieved by fitting for the parameters $a$
and $b$. It is found that for the white light the above representation
of the wide-band magnitudes holds with an accuracy of 4 per cent for
$W-$magnitudes and 2 per cent, for $P-$magnitudes. Unfortunately, for a
polarizer that we used in the observations there was a long-wavelength
interval with unspecified transmission efficiency function. Simulating
polarizer transmission function in different ways and folding it with
different spectra (Vega's, $\propto \nu^{-1}$, and $\propto \nu^0$), we
have estimated the maximum additional uncertainty of $\sim 20$ per cent.
Thus we add 20 per cent of uncertainty when we  convert $P$ to absolute
fluxes.

Observed magnitude $W$ of an investigated source is converted
into the absolute flux values using the Pogson equation
\begin{equation}
F_\mathrm{GRB} ^{W} = F_\mathrm{o}^{W} \times  10^{-0.4\,W} \, ,
\label{eq:pogson}
\end{equation}
where $F_\mathrm{o}^{W}$ is the calculated Vega flux
in the CCD  spectral band: 
\begin{equation}
F_\mathrm{o}^W = \int F_\mathrm{Vega}(\lambda) \, W(\lambda) \, \mbox{d}\lambda = 1.33\times 10^{-5}
\mathrm{erg}\, \mathrm{cm}^{-2}\, \mathrm{s}^{-1}\, .
\label{eq:Vega_ccd_flux}
\end{equation}
In the same way, magnitudes $P$  are converted to flux values using
eq. (\ref{eq:pogson}) with the zero-magnitude flux
\begin{equation}
F_\mathrm{o}^P =\int F_\mathrm{Vega}(\lambda) \, P(\lambda) \,\mbox{d}\lambda = 1.16\times 10^{-5}
\mathrm{erg}\, \mathrm{cm}^{-2}\, \mathrm{s}^{-1}\, .
\label{eq:Vega_ccd_pol_flux}
\end{equation}
In the above expressions, $ W(\lambda)$ and $P(\lambda)$ are,
respectively, the normalized CCD-response function and its normalized
convolution with the transmission efficiency of the polarizer (see the
right panel of Fig.~\ref{fig:calibration}). Of course, if one observes a
source with partly (linearly) polarized light through such polarization
filter, one needs a full information on the degree of polarization and
its angle to derive the total flux. For an unpolarized source, or for a
source with equal signals from two polarizers (the case of
GRB\,100906A), we can safely use formula (\ref{eq:Vega_ccd_pol_flux}).
For details we refer to Appendix.
\begin{table*}
 \centering \begin{minipage}{140mm} \caption{Stars from SDSS-DR7
catalogue used as standard stars for GRB\,100901A. Columns 3--7 contain
SDSS $ugriz$ magnitudes (Vega system). Column 8 contains average
unfiltered $W$-band MASTER magnitude. Full table can be found online in
the electronic issue. }
\label{tab:st_stars_100901}
  \begin{tabular}{@{}cccccccccc@{}}
  \hline
Source                 &
SDSS                   &    
$u$                       &
$g$                       & 
$r$                       &
$i$                       &
$z$                       &
$W$ 			&
RA\,J2000                 &
DEC\,J2000                 \\
                       ID &
name                   &    
(mag)                       &
(mag)                       & 
(mag)                       &
(mag)                       &
(mag)                       &
(mag)                       &
(d)                 &
(d)                 \\
\hline
1	&	J014906.68+224024.7	&	19.246	&	17.05	&	15.96	&	15.462	&	15.161	&	15.95	&	27.277834	&	22.673535	\\
2	&	J014904.39+224045.3	&	15.841	&	14.609	&	14.143	&	13.953	&	13.897	&	14.21	&	27.26829	&	22.679256	\\
\hline
\end{tabular}
\end{minipage}
\end{table*}
\begin{table*}
 \centering
 \begin{minipage}{140mm}
  \caption{Standard stars used for the photometry of GRB\,100906A.
Column 3 contains average unfiltered $P$-band MASTER magnitude measured 
with one of polarization filters.  Full  table can
be found online in the electronic issue.}
\label{tab:st_stars_100906}
  \begin{tabular}{@{}ccccccc@{}}
  \hline
Source&USNO-A2.0&$P_{\updownarrow}$&Standard deviation&
$\langle P_{\updownarrow}\rangle-\langle P_{\leftrightarrow}\rangle$&RA\,J2000&DEC\,J2000\\
ID  & name &     (mag)   &      of light curve (mag)     &(mag) &(d)&(d) \\
\hline
(1)   & (GRB\,100906A)  & \dots &1.14  &0.01    & 28.68413  &55.63050       \\
2   & 1425-02624804 & 15.70 & 0.04 &  -0.04  & 28.69122 & 55.63834       \\
\hline    
\end{tabular}
\end{minipage}
\end{table*}

To obtain GRB's flux density at the wavelength of the CCD-response
maximum, $5500$~\AA, we divide flux $F_\mathrm{GRB}^{W}$ or
$F_\mathrm{GRB}^{P}$ by an effective frequency interval $\Delta
\nu_\mathrm{eff}$ of the corresponding response function. Naturally, the
value of $\Delta \nu_\mathrm{eff}$ depends on a particular spectral
distribution. The effective frequency interval of the CCD-response
function is \hbox{$\sim 3.9\times10^{14}$~Hz}, calculated as the mean of
values $\Delta \nu_\mathrm{eff}$ for the Vega spectrum, white light, and
power-law spectrum with $-1$ slope, all falling inside a 15 per cent
uncertainty interval.
 The effective frequency interval of the CCD plus polarizer, calculated
in the same way, is $\sim 3.2\times10^{14}$~Hz with a 10 per cent
uncertainty interval.
 Bearing in mind that a real spectrum might be something different from
the three considered in such simple analysis, we assume approximately a
20 per cent accuracy for converting from flux units to
spectral flux density units.

\subsection{{\em Swift} BAT and XRT light curves\label{section:bat_xrt}}

In order to build flux light curves observed by {\em
Swift}/BAT~(Figs.~\ref{fig:lc100901a_all} and \ref{fig:lc100906a_all}), 
standard sets of GRB products are obtained using the ftool
{\tt batgrbproduct} within HEASOFT v6.10, and the
count light curves in 15-350~keV are retrieved.
 The $15-150~$keV spectra averaged over the total-fluence time interval
are analysed in XSPEC \citep{arnaud1996} using the power-law
approximation. The derived spectral parameters are used to calculate
$15-350$~keV model flux and to obtain counts to flux conversion factors
in the $15-350$~keV range. Thus we apply constant conversion
factors $1.07\times 10^{-6}$ erg cm$^{-2}$ counts$^{-1}$ for
GRB\,100901A and $9.17\times 10^{-7}$ erg cm$^{-2}$ counts$^{-1}$ for
GRB\,100906A.
 
The {\em Swift}/XRT 0.3-10~keV unabsorbed light curves are available
from the {\em Swift} Burst Analyser. These light curves are produced
with information on spectral evolution involved~\citep{evans_etal2010}.
The late-time spectra of GRBs, available from the {\em Swift} Spectra
Repository, are the source of estimates of the column densities for the
soft X-ray absorption.

\subsection{Spectral fitting}
\label{section:fitting}

In order to build simultaneous wide-energy spectra, we select time
intervals, which are covered by the optical observations. The prompt
stage of GRB\,100901A (before $T_{90}$) contains three suitable MASTER
exposure intervals: $t-T0=193-233$, $312-372$, and $387-467$~s, where
$T0$ is the trigger time. There are two MASTER time intervals during the
prompt stage of GRB\,100906A overlapping with the observations by {\em
Swift}/XRT, $73.8-83.8$~s and $105-125$~s (or very close, in the first
case).

 We download the X-ray time-sliced spectra, constructed for the time
intervals of the optical exposures, from the {\em Swift} spectra
repository~\citep{evans_etal2009}. For GRB\,100906A we also use the
results obtained by the Konus--{\em Wind}~\citep{gcn11251}. BAT spectra
and response matrices are prepared as described in the Data Analysis
Guidelines\footnote{
http://swift.gsfc.nasa.gov/docs/swift/analysis/threads\\/batgrbproductthread.html}.
They are extracted for the selected time intervals and corrected using
the tasks {\tt batbinevt}, {\tt batupdatephakw}, {\tt batphasyserr}, and
a response matrix generated by the {\tt batdrmgen} task. Then the
spectra are fitted in XSPEC v12.6 using models {\tt powerlaw} (power
law), {\tt cutoffpl} (cut-off power law), or {\tt grbm} (Band function),
in combination with absorption components. When simultaneous fits for
the $0.3-10$~keV and $15-150$~keV and/or optical data are performed, XRT
spectra are rebinned using {\tt grppha} to ensure that there are at
least 20 counts per bin, which is needed for joint minimization of
$\chi^2$.

To account for the soft X-ray absorption, we use XSPEC model component
{\tt phabs} (Galactic absorption) and {\tt zphabs} (absorption in the
host galaxy). When we fit the parameters of the optical extinction in
the GRB host, we use the XSPEC model {\tt zdust}, which is the
extinction by the dust grains. In the fits, whose results are presented
in the tables in the next section, we assume the Small Magellanic Cloud
(SMC) type of the host galaxy extinction curve and the
total-to-selective extinction $R_V=2.93$~\citep{Pei1992} following
results of the studies by \citet{Jensen_etal2001,Schady_etal2010}. We
discuss variations due to the different extinction laws in the text.
Correction for the reddening in the Galaxy is done beforehand, using the
values calculated using the NED Galactic Extinction
calculator\footnote{http://ned.ipac.caltech.edu/forms/calculator.html}
based on a work of \citet{Schlegel_etal1998}.

To analyse the late-time spectra, we consider the optical observations
after $10\,T_{90}$. First, we check that big variations of X-ray flux
are absent in a time interval: we request that the flux density at 
$10$~keV varies less by a factor of 3. The X-ray data is rebinned to a
minimum of 10 (20, if possible) counts per bin (by {\tt grppha}). The
optical data is averaged over the time interval and transformed by the
task {\tt flx2xsp}. We check that the number of degrees of freedom in
the fit by the XSPEC model {\tt phabs}$*${\tt zphabs}$*${\tt
zdust}$*${\tt powerlaw} is more than 5. Thus selected time intervals are
depicted by rectangles in Figs.~\ref{fig:lc100901a_all},
\ref{fig:lc100906a_all}, and \ref{fig:break_time}.

\section{Results}
\label{section:results}

\subsection{Spectra of GRB\,100901A and  GRB\,100906A before $T_{90}$}
\label{s.lc_spectra}

\begin{figure*}
\includegraphics[width=0.45\textwidth]{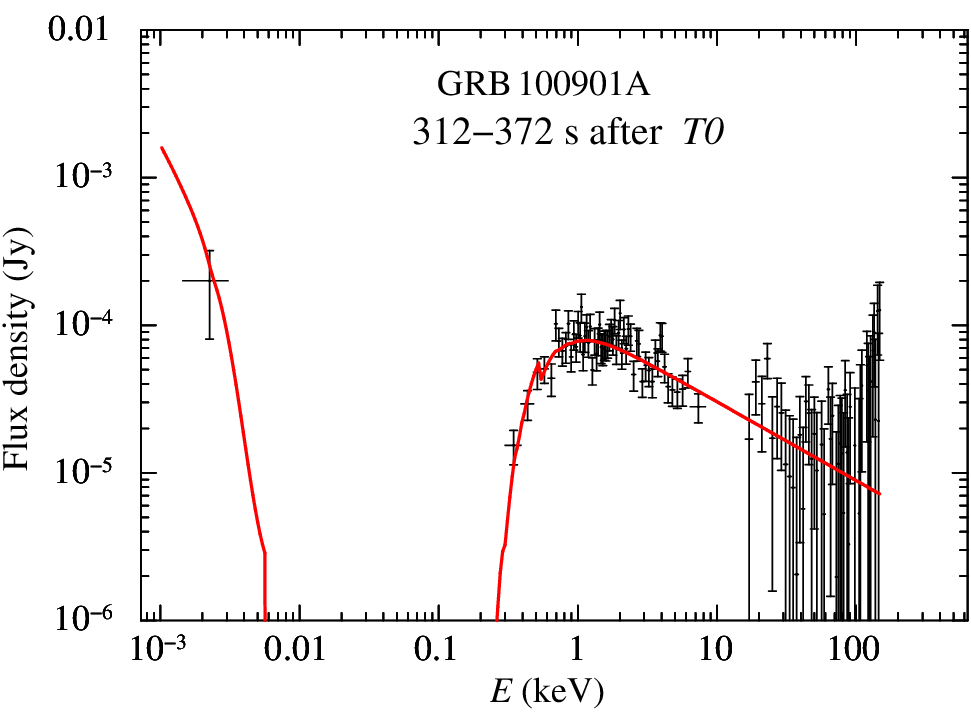}
\hskip 5mm
\includegraphics[width=0.45\textwidth]{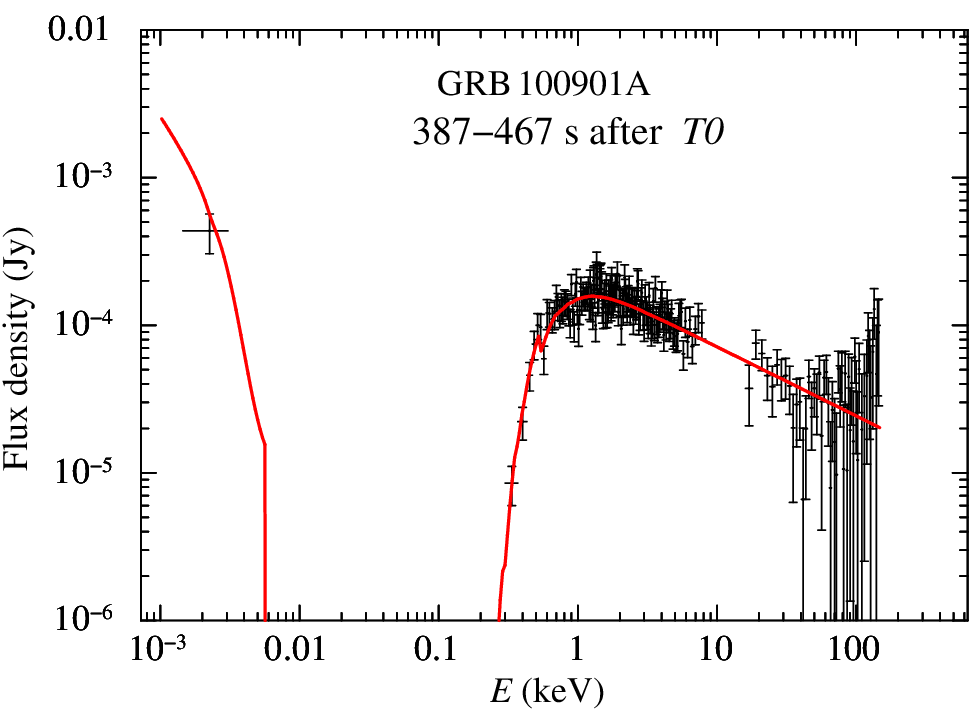}
\caption{ Spectrum of GRB\,100901A for two time intervals at $t\la
T_{90}$. Optical flux density obtained by MASTER, corrected for the
Galactic extinction $A_V=0.327$ \citep[NED; ][]{Schlegel_etal1998}, is
shown by the single left point, whose horizontal bar corresponds to the
MASTER unfiltered effective frequency interval. Spectra in 0.3--10 and
15--150~keV are made with the {\em Swift} BAT and XRT data. 
Best-fitting absorbed power laws are shown by the red lines. Their
spectral parameters are described in Table~\ref{tab:spectral_indexes} as
Fit 100901.2 for $312-372$~s and Fit 100901.3, for $387-467$~s).
\label{fig:spectrum100901}}
\end{figure*}

For the three early spectra of
GRB\,100901A~(Fig.~\ref{fig:spectrum100901}), fits with
absorbed power-law are satisfactory (Fits 100901.1, 100901.2, and 100901.3
in Table~\ref{tab:spectral_indexes}). Other spectral models, cut-off
power law and the Band function, do not improve the fitting statistics.

\begin{figure*}
\includegraphics[width=0.45\textwidth]{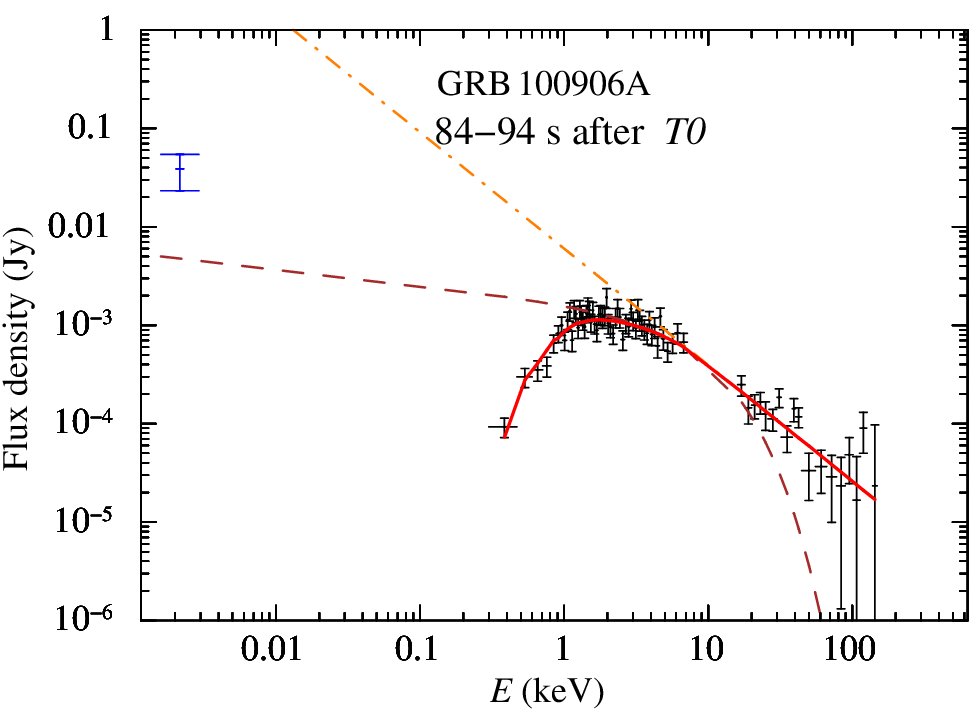}
\hskip 5mm
\includegraphics[width=0.45\textwidth]{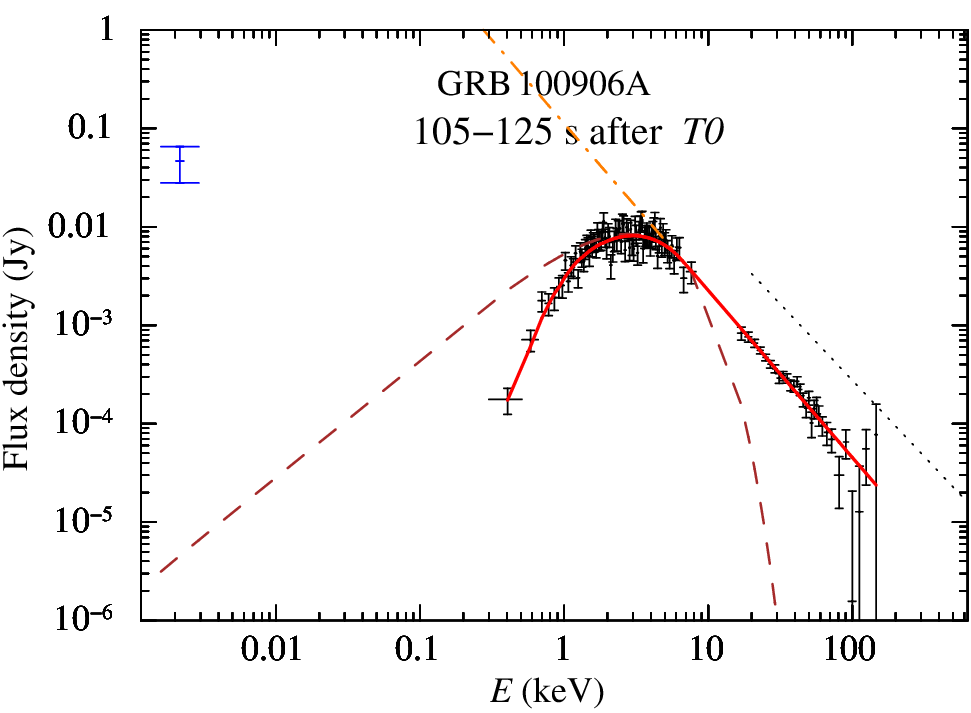}
 \caption{ Spectrum of GRB\,100906A for two time intervals at $t\la
T_{90}$.
Optical points are corrected for the Galactic extinction
 $A_V=1.194$ \citep[NED; ][]{Schlegel_etal1998}.
In the left panel, we use the MASTER observation at 73.8--83.8~s. 
Best-fitting absorbed Band functions are shown by the red 
lines (Fit 100906.1 and 100906.2 in
Table~\ref{tab:spectral_indexes}). No attempt has been made to estimate
the optical extinction in the GRB host galaxy.
 The brown dashed line depicts the
unabsorbed low-energy part, and the orange dot-dashed line, the
unabsorbed high-energy part. Dotted power law represents observations of
Konus--{\em Wind} from 98.304 to 122.880~s in 20~keV-2~MeV with a
correct slope and a roughly estimated flux. 
\label{fig:spectrum100906}}
\end{figure*}

For GRB\,100906A, we do not include the host-galaxy absorption model
component, as its spectral distributions from the optical to X-rays
(Fig.~\ref{fig:spectrum100906}) cannot apparently be fitted by a single
power law. Generally, for a variable complex spectrum of an early GRB
emission, it is impossible to estimate the extinction in the optical
band, as one can use no \hbox{a-priori} information about the spectral
distribution. Table~\ref{tab:spectral_indexes_06} contains the
parameters of the fits by the Band model, together with the resulting
values of $\chi^2$ for the alternative models with fewer free
parameters, which give worse results. We also calculate the slopes of
the photon spectrum between the MASTER points and the best-fit 3~keV
flux densities (column 3 in Table~\ref{tab:spectral_indexes_06}). We
remind that the optical points are corrected for the Galactic
extinction.

Apparently, the peak of the spectral energy distribution (SED) of
GRB\,100906A is located somewhere around 0.3--30~keV at times 80--120~s
and moves gradually to the soft energies. The SED-peak energy
$E_\mathrm{peak}$ for the Band spectrum is $(-\alpha_\mathrm{B}+2)\times
E_0$, where $\alpha_\mathrm{B}$ is the photon index of the lower-energy
part of the Band function, and $E_0$ is the characteristic energy for
the Band function (Table~\ref{tab:spectral_indexes_06}). We obtain that
$E_\mathrm{peak}$ shifts from about 30 to 4~keV between the two spectral
distributions presented in Fig.~\ref{fig:spectrum100906}. The higher
energy part of the spectrum of GRB\,100906A ($>30$~keV) can be fitted by
a single power law. This is consistent with the Konus--{\em Wind}
results (Fig.~\ref{fig:spectrum100906}, dotted line). According to
\citet{gcn11251}, the spectrum of the second bursting episode of
GRB\,100906A observed by Konus--{\em Wind} in $\gamma$-rays from
$98.304$ to $122.880$~s is best fitted in the 20~keV--2~MeV range by a
power-law model with the photon index $2.55_{-0.2}^{+0.25}$.

\begin{table*}
\caption{Spectral parameters for GRB\,100901A at the
prompt stage resulted from the 
simultaneous fitting of the absorbed power law
in XSPEC ({\tt phabs}$*${\tt zphabs}$*${\tt zdust}$*${\tt powerlaw})  to
the MASTER, XRT, and BAT data~(Fig.~\ref{fig:spectrum100901}).
Galactic absorption column density
is  fixed at $N_\mathrm{H} = 7.1\times 10^{20}
\mathrm{cm}^{-2}$~\citep{kalberla_etal2005}.
Assumed host galaxy extinction curve is SMC 
type and the total-to-selective extinction is  $R_V=2.93$.
 The 4th and 5th columns give the
fitted intrinsic absorption column density and  optical extinction
at $z=1.408$, respectively.
 }
\label{tab:spectral_indexes}
  \begin{tabular}{@{}cccccccc@{}}
\hline
Fit ID &$t-T0$  (s) & 
Photon index &
$N_\mathrm{H}^{\mathrm{int}}$  $(10^{22}\, $cm$^{-2}$)  &
$A_V^{\mathrm{int}}$            &$ \chi^2/\mathrm{dof}$ \\
\hline
100901.1&
$ 193-233$    &
$1.60\pm 0.08$ &
$0.6\pm 0.2$ & $1.1_{-0.3}^{+0.6}$&
 73.4/94 \\
100901.2&
$ 312-372$    &
$1.53\pm 0.05$ &
$0.4\pm 0.1$ & $0.95_{-0.2}^{+0.4}$&
 138/134 \\
100901.3 &
$387-467$    &
$1.49\pm 0.02$ &
$0.55\pm 0.07$  & $0.75_\pm 0.1$ &
216/219\\
\hline
\end{tabular}
\end{table*}
\begin{table*}
\caption{
Spectral parameters for GRB\,100906A at the prompt stage resulted from the 
fitting of the absorbed Band function
in XSPEC ({\tt phabs}$*${\tt zphabs}$*${\tt grbm})  to
the  XRT and BAT data~(Fig.~\ref{fig:spectrum100906}).
Galactic absorption column density is fixed at $N_\mathrm{H} =2.2\times 10^{21}
\mathrm{cm}^{-2}$~\citep{kalberla_etal2005}.
Column 3 gives the photon index between the optical and modeled 3~keV
flux density;
next columns:
the photon indices and  characteristic energy of the Band function,
intrinsic absorption at $z=1.727$. In the last three columns statistics $\chi^2$ of the fits
shown in Fig.~\ref{fig:spectrum100906} are typed in bold, statistics of
fits with the alternative models are typed in normal font. }
\label{tab:spectral_indexes_06}
  \begin{tabular}{@{}ccccccccccc@{}}
\hline
Fit  &$t-T0$ & MASTER-XRT    &
\multicolumn{2}{|c|}{Photon indices}    & 
$E_0$   &
$N_\mathrm{H}^{\mathrm{int}}$             &$ \chi^2/\mathrm{dof}$
& $\chi^2/\mathrm{dof}$ & $\chi^2/\mathrm{dof}$ \\
  ID & (s)                   & 2.2~eV--3 keV   &
\multicolumn{2}{|c|}{of Band function}&  (keV)         &
$(10^{22}\, $cm$^{-2}$) & Band    & power law &
cut-off pow \\
\hline
100906.1         &
$84-94^\mathrm{\dagger} $     &
$1.50 \pm 0.06  $       &
$1.2\pm 0.2$ & $2.18_{-0.07}^{+0.08}$&
$9^{+7}_{-4}$  &$ 0.1_{-0.1}^{+0.3}$ & 
\bf{112/110}  &
{141/112 }            & {121/111} \\
100906.2         &
$105-125 $        &
$1.24\pm 0.06 $      &
 $-0.22^{+0.24}_{-0.07}$ &  $2.70 \pm 0.03$  &
$2.3^{+0.4}_{-0.1} $ & $0_{-0.0}^{+0.23}$&   
\bf{149/139 }& 
{314/141}        & {254/140}  \\
\hline
\multicolumn{10}{p{16.5cm}}{%
$^\mathrm{\dagger}$ MASTER observation was carried out in an slightly earlier
 time interval, 73.8--83.8~s.}\\
\end{tabular}
\end{table*}

\subsection{Host galaxy extinction  of GRB\,100901A}
\label{section:ext01}

\begin{table*}
\caption{Spectral parameters for GRB\,100901A at late time
intervals. XSPEC fitting of the absorbed power law 
to the MASTER and XRT/{\em Swift} data
is performed. Frozen fit parameters
are the same as for the fits in Table~\ref{tab:spectral_indexes}. 
 }
\label{tab:spectral_indexes_01_late}
  \begin{tabular}{@{}cccccccc@{}}
\hline
Fit ID  &$t-T0$ (s) & 
Photon index &
$N_\mathrm{H}^{\mathrm{int}}$  ($10^{22}\, $cm$^{-2}$)  &
$A_V^{\mathrm{int}}$            &$ \chi^2/\mathrm{dof}$ \\
\hline
100901.4&
$ 5000-6500$    &
$2.1\pm 0.1$ &
$0.3\pm 0.1$ & 
$0.8_{-0.4}^{+0.5}$&    
15.8/14 \\
100901.5&
$ 15000-17500$    &
$2.0\pm 0.1$ &
$0.1_{-0.1}^{+0.2}$ & 
$0.2^{+0.3}_{-0.2}$&    
38.6/38 \\
100901.6&
$ 27000-28500$    &
$2.1\pm 0.1$ &
$0.3\pm 0.2$ & 
$0.4^{+0.4}_{-0.3}$&    
36.6/35 \\
100901.7 &
$102800-121500$    &
$2.10_{-0.11}^{+0.01}$ &
$0.28 \pm 0.25 $  &  
$0.5_{-0.2}^{+0.8}$ & 
13.0/10\\
\hline
\end{tabular}
\end{table*}

We fit several late-time spectra comprised of the XRT/{\em Swift} data
and simultaneous optical points available from MASTER. The optical
points are the time-averaged unfiltered $W$ magnitudes as follows:
$18.05\pm 0.03$, $17.78\pm 0.02, $ $17.61 \pm 0.04$, and $19.30\pm 0.02$~mag, 
corrected for the Galactic extinction $A_V=0.327$
\citep[NED; ][]{Schlegel_etal1998},
successively
for the fits in Table~\ref{tab:spectral_indexes_01_late}. As well as for
the prompt spectra, we can approximate the late-time spectra of
GRB\,100901A by an absorbed power-law model.

Averaging results of the  late-time fits 
(Table~\ref{tab:spectral_indexes_01_late}), we obtain 
$A_V^{\mathrm{int}} = 0.4\pm
0.2$  and $N_\mathrm{H}^{\mathrm{int}}= (2.2\pm 0.7) \times 10^{21} \mathrm{cm}
^{-2}$.
Averaging results of the three early fits 
(Table~\ref{tab:spectral_indexes}), we obtain  $A_V^{\mathrm{int}} = 0.8\pm 0.1$ 
 and $N_\mathrm{H}^{\mathrm{int}} = (5.1 \pm 0.6)\times 10^{21} \mathrm{cm}
^{-2}$. 
There is an apparent  change in values of $N_\mathrm{H}^{\mathrm{int}}$ and
$A_V^{\mathrm{int}}$ between the prompt-time and late-time fit parameters. 
It is likely that the prompt-time value of $N_\mathrm{H}^{\mathrm{int}}$ is biased
because of the spectral evolution and, probably, because the prompt
spectral shape is not precisely power-law in the range from the optical to
X-rays.
\cite{Butler-Kocevski2007} and \cite{evans_etal2010}  argue
that absorption derived from early-time
XRT data, when a strong spectral evolution is present, can be misleading.

The late-time spectrum compiled at the XRT/{\em Swift} spectrum
repository (UK {\em Swift} Science Data Centre) for 4000--33229~s yields
the value of the intrinsic $N_\mathrm{H}^{\mathrm{int}}=(2.9\pm
0.9)\times 10^{21} \mathrm{cm} ^{-2}$ (90 per cent uncertainty interval), which
is in a very good agreement with our fit results for the late-time
sliced spectra. This indicates that a power law approximation for the
X-ray spectrum of GRB\,100901a is indeed a good approximation for any
late time interval.

We have also tried other types of extinction laws in the GRB host. In
addition to the fits, presented in 
Table~\ref{tab:spectral_indexes_01_late}, we performed fits with
different extinction laws in the host galaxy: that of the LMC
(Large Magellanic Cloud) and MW (Milky Way). For them we applied
different values of $R_V$~\citep[][see below]{Pei1992}. The quality of
the fits remains the same and the values of $N_\mathrm{H}$ as well, 
because the choice of the optical extinction does not affect strongly 
the amount of the soft X-rays absorption. We have found that depending
on the extinction law applied, for three late-time intervals, the
average total extinction $A_V^\mathrm{int} = R_V \,E(B-V)^\mathrm{int}$ 
varies as: 
$\langle A_V^\mathrm{int} \rangle = 0.4\pm 0.2$ (SMC, $R_V =2.93$), 
$\langle A_V^\mathrm{int} \rangle = 0.5\pm 0.2$ (LMC, $R_V =
3.16$), 
and $\langle A_V^\mathrm{int} \rangle = 0.7\pm 0.2$ (MW, $R_V =3.08$).

Basing on the `late-time'
values of the optical extinction in the GRB host we calculate the 
ratio $N_\mathrm{H}^\mathrm{int}/A_V^\mathrm{int}$.
 It equals
$(5\pm 2)\times 10^{21} \mathrm{cm} ^{-2}$ 
(for the SMC type of the extinction in the GRB host), 
$(4\pm 2)\times 10^{21} \mathrm{cm} ^{-2}$ (LMC) and 
$(3\pm 1)\times 10^{21} \mathrm{cm} ^{-2}$ (MW). 
These are
not far from the ratios empirically found for the Milky Way and
Magellanic Clouds: $\sim 4\times 10^{21}$ (SMC), $\sim 3.5\times
10^{21}$ (LMC), and $\sim 2\times 10^{21}\mathrm{cm}^{-2}$ (MW)
\citep[e.g.,][]{Schady_etal2010}. Thus, independently of the extinction
law used, the resulting metal-to-dust ratio in the host of GRB\,100901A
is comparable to the level measured for the Milky Way or Magellanic
Clouds.

\subsection{Host galaxy extinction  of GRB\,100906A}
\label{section:ext06}
\begin{table*}
\caption{Spectral parameters for GRB\,100906A at several late time
intervals (Figs.~\ref{fig:spec100906_27000-31000} and 
\ref{fig:sed100906_50000-54000}). XSPEC fitting of the absorbed power law
or broken power law to the MASTER, OSN, and XRT/{\em Swift} data
is performed.
Galactic absorption column density
is  $N_\mathrm{H} = 2.21\times 10^{21}
\mathrm{cm}^{-2}$~\citep{kalberla_etal2005}.
The host galaxy extinction curve is that of SMC (Small Magellanic
Cloud) and the total-to-selective extinction is taken $R_V=2.93$.
The columns designation is the same as for
Table~\ref{tab:spectral_indexes_01_late} with an addition of the lower 
photon index (left) and the break energy in eV. }
\label{tab:spectral_indexes_06_late}
  \begin{tabular}{@{}cccccccccc@{}}
\hline
Fit ID &$t-T0$ (s) & 
Photon index & $E_\mathrm{break}$ (eV) & Photon index &
$N_\mathrm{H}^{\mathrm{int}}$  ($10^{22}\, $cm$^{-2}$)  &
$A_V^{\mathrm{int}}$            &$ \chi^2/\mathrm{dof}$ \\
\hline
100906.3&
$ 6000-7000$    &
 \dots &\dots & $1.96\pm 0.05$ &
$0.4_{-0.2}^{+0.3}$ & $<0.4$
&  36.6/42 \\
&
$ 10500-11000$    &
 \dots &\dots & $1.9\pm 0.1$ &
&&   \\
&
$ 28000-30400$    &
 \dots &\dots & $2.01\pm 0.02$ &
&&  \\
100906.4 &
$45000-65000$    &
$-1.3\pm 0.7 $ & $2.5\pm 0.3$ &
$2.34\pm {0.04}$ &
$0.25_{-0.25}^{+0.55}$  & $0.35\pm 0.09$ &
9.78/10\\
\hline
\end{tabular}
\end{table*}

\begin{figure}
\includegraphics[width=0.45\textwidth]{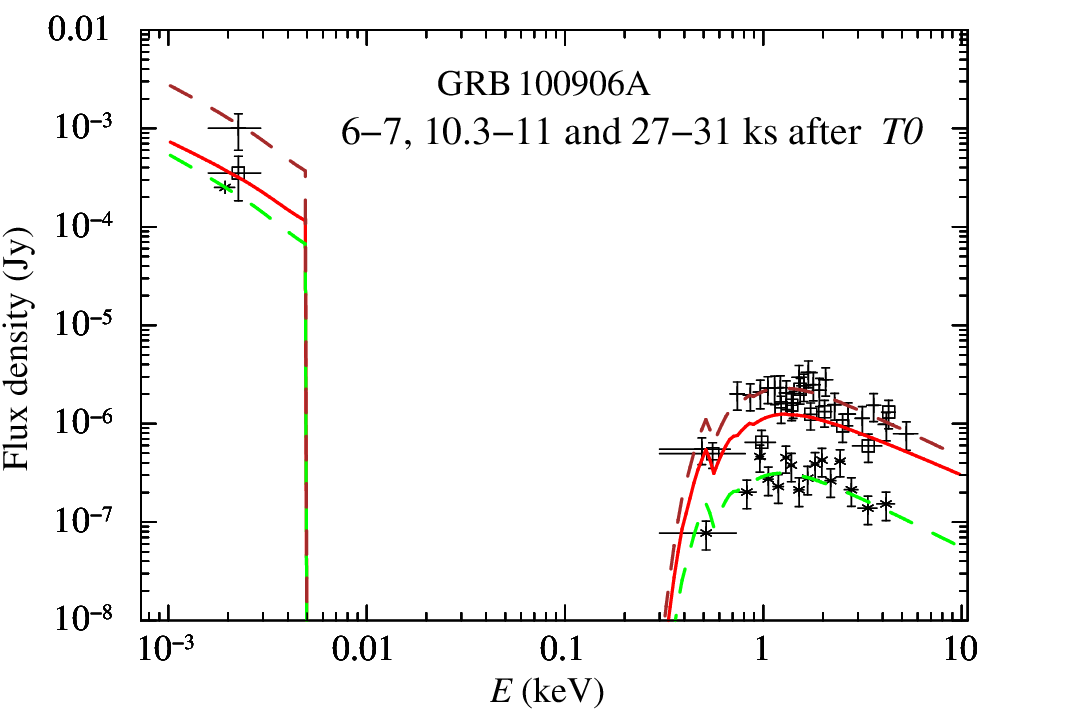}
\caption{
\label{fig:spec100906_27000-31000}
Spectral distributions of GRB\,100906A at three time intervals:
6000-7000, 10500--11000 and 28000-30400~s (points, squares and crosses 
with bars, respectively).
The optical data are: the MASTER $P=17.54\pm 0.05$ for $6000-7000$~s,
$P=18.7\pm 0.3$ for
$10500-11000$~s and OSN $R=18.70\pm 0.02$ for for $28000-30400$~s,
additionally corrected for the Galactic extinction 
$A_V=1.194$ and $A_R=0.963$~\citep[NED; ][]{Schlegel_etal1998}.
Lines show the best-fitting absorbed power laws, whose 
parameters are listed as Fit 100906.3 in 
Table~\ref{tab:spectral_indexes_06_late}.
}
\end{figure}

\begin{figure}
\includegraphics[width=0.45\textwidth]{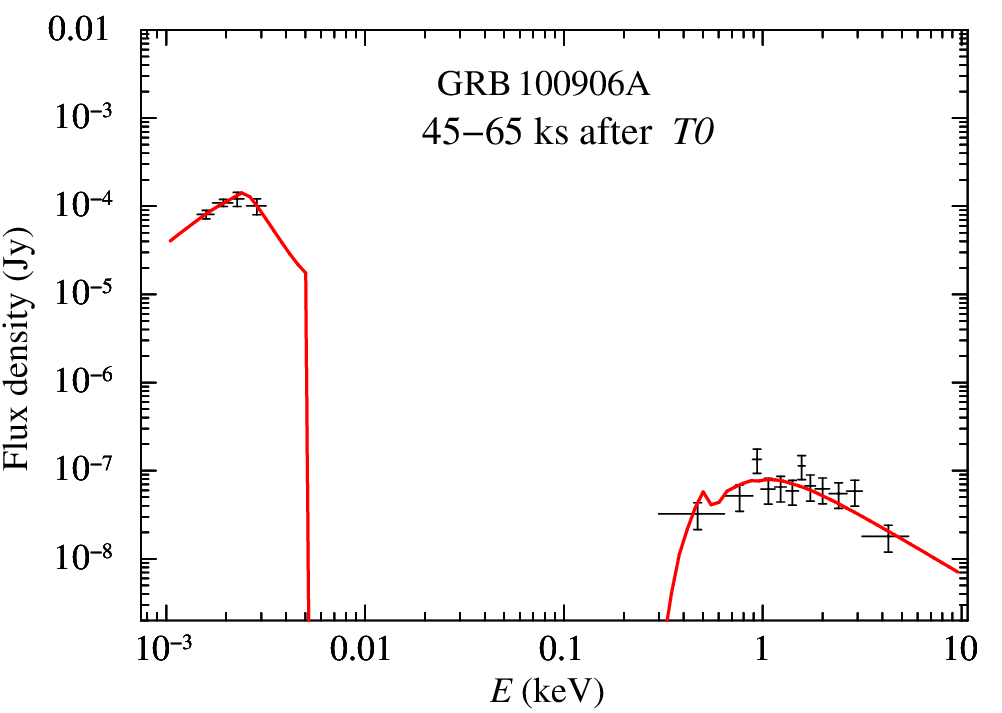}
\caption{
Spectrum of GRB\,100906A compiled from the XRT data  in the interval
45--65~ks
and optical $BVIR$ observations obtained by  OSN around 14~h
after the trigger (in the interval 50.9--54.0~ks) 
and corrected for the Galactic extinction.
Solid line is the best-fitting absorbed broken power law model, whose 
parameters are listed as Fit 100906.4 
in Table~\ref{tab:spectral_indexes_06_late}.
\label{fig:sed100906_50000-54000}}
\end{figure}

 Three late-time spectral distributions are shown in
Figs.~\ref{fig:spec100906_27000-31000} and 
\ref{fig:sed100906_50000-54000}. The parameters of the fits are listed
in Table~\ref{tab:spectral_indexes_06_late}. The optical points are the
time-averaged unfiltered $P$ magnitudes by MASTER: $17.54\pm 0.05$ for
$6000-7000$~s, $18.7\pm 0.3$ for $10500-11000$~s; OSN $R=18.70\pm 0.02$
for for $28000-30400$~s. The optical point in the $R$-band at
50000--54000~s is obtained by averaging the OSN data ($R=19.7\pm 0.3$)
and $V, B, I$ points can be found in Table~\ref{tab:OSN100906}. Before
the fitting, the optical data are corrected for the Galactic extinction
calculated in the NED Extragalactic Calculator: $A_V=1.194$,
$A_B=1.554$, $A_R=0.963$, and $A_I=0.699$ ~\citep[NED;
][]{Schlegel_etal1998}.

One can see that the optical data at $\sim
14$~h (Fig.~\ref{fig:sed100906_50000-54000}) is best fitted by an
absorbed broken power-law with a peak in the optical range. However, we
cannot consider this peak robust  because the optical points
are non-simultaneous and there is a strong variation observed in the
$R$-band around 14~h (see Fig.~\ref{fig:lc100906a_opt_OSN}).

The parameters of the Fit 100906.3 (the first three lines in
Table~\ref{tab:spectral_indexes_06_late}) are obtained by simultaneously
fitting the data for the three time intervals with the spectral slope
and normalization left independent and tied $E(B-V)^{\mathrm{int}}$
(that is, $A_V$) and $N_\mathrm{H}^\mathrm{int}$. This was done in an
attempt to increase statistics to obtain positive 1$\sigma$ interval for
$A_V^\mathrm{int}$. Nevertheless, we can report only the 2$\sigma$ upper
limit. Setting other types of the extinction law in the host galaxy 
does not change its value.

The spectrum of GRB\,100906A at 10--50~ks interval 
 from the XRT spectrum repository (at the UK {\em Swift} Science Data
Centre) yields the value of intrinsic
\hbox{$N_\mathrm{H}^\mathrm{int}=(8.5\pm 3)\times 10^{21}$} cm$^{-2}$
(with 90 per cent uncertainty interval). The values of
$N_\mathrm{H}^\mathrm{int}$ in Table~\ref{tab:spectral_indexes_06_late}
with 1$\sigma$ errors are consistent with this value.
If we take the lower 2$\sigma$ limit of the XRT repository 
value, $N_\mathrm{H}^\mathrm{int}=4.8\times 10^{21}$ cm$^{-2}$,
and the upper 2$\sigma$ limit on $A_V^{\mathrm{int}} $ obtained for different
extinction laws,
we arrive at the lower 2$\sigma$ limit   on
$N_\mathrm{H}^\mathrm{int}/A_V^\mathrm{int}\approx 1.1\times 10^{22}$.
This limit is at least twice the typical ratios, empirically defined for the 
Milky Way and Magellanic Clouds (see \S~\ref{section:ext06}). 
This  ratio 
is consistent with the results of \citet{Schady_etal2010} who studied 28
GRBs and found values of $N_\mathrm{H}^\mathrm{int}/A_V^\mathrm{int}$
typically higher than those for the Milky Way and Magellanic Clouds.

We conclude that for GRB\,100906A our fitting provides only the upper
limit on $A_V^\mathrm{int}$, and that there are indications that this GRB has 
high $N_\mathrm{H}^\mathrm{int}/A_V^\mathrm{int}$ ratio, which can be due to a dust destruction or
 the intrinsics properties of the host galaxy.

\subsection{Jet break time of GRB\,100906A}
\label{section:jetbreak}

\begin{table*}
 \centering \caption{Parameters of the time fits to the GRB\,100906A
light curves, where slope $\alpha$ is quoted in $F_\nu \propto
t^{-\alpha} \nu^{-\beta}$. The first three fits are done with the broken
power-laws, and the last one, with the power law. Also given are the
spectral indices in the XRT spectral range
$\langle\beta_\mathrm{XRT}\rangle$, retrieved from the {\em Swift} Burst
Analyser and averaged over the relevant time intervals. The time
intervals are the two parts of the whole time span (2nd column), divided
at the break time (5th column).}
\label{tab:100906_breaks}
  \begin{tabular}{@{}cccccccc@{}}
  \hline
Light curve& Time  (ks) & 
\multicolumn{2}{|c|}{ before break }&
$t_\mathrm{break}$ (ks)  &
\multicolumn{2}{|c|}{after break } &
 $\chi^2/\mathrm{dof}$\\
&   &$\alpha$ & $\langle\beta_\mathrm{XRT}\rangle$ & & $\alpha$ &  $\langle\beta_\mathrm{XRT}\rangle$  &\\
\hline
$3-10$~keV & $10-300$ &  
$ 1.68 \pm 0.02 $ & $1.0\pm 0.3$& 
$51_{-3}^{+4}$ &
$2.9_{-0.3}^{+0.1}$ &$1.1\pm 0.2$  &  145/78 \\
$R$  (OSN) & $28-54$ & 
$1.36\pm 0.05$ & 
$1.1\pm 0.2$ &
 $51.9_{-0.2}^{+0.5}$  &
$10^{+13}_{-3}$ &
\dots &
 56.0/55 \\
 $R$ (OSN$+$NOT)  & $28-310$ & 
$0.14 \pm 0.02$ & 
$1.1\pm 0.2$  &
$34.9_{-0.2}^{+0.5}$  & 
$2.17_{-0.04}^{+0.03}$ &
$1.0\pm 0.2$ &
135/58 \\
1-3~eV (MASTER) & $0.18-2.52$ & 
$1.08 \pm 0.02$ & $1.3 \pm 0.4$ & 
$\dots$ &
$\dots$ & 
$\dots$ &
13.9/12\\
\hline
\end{tabular}
\end{table*}

We construct broken power-law fits to the light curves of GRB\,100906A
after 10~ks. Table~\ref{tab:100906_breaks} presents the parameters of
the fits. A break on the X-ray light curve is found around $\sim 50$~ks.
The parameters of the $0.3-10$~keV light curve can be reconciled with
the ``closure relations'' between the temporal and spectral indices
($\alpha$ and $\beta$ in $F_\nu \propto t^{-\alpha}\nu^{-\beta}$) in the
framework of the fireball model in the following way (see table 1 in
\citet{racusin_etal2009}, references therein and reviews of 
\citealt{meszaros2002,zhang-meszaros2004,piran2005}). Before the break,
for the ISM environment and the `slow cooling' regime and using 
$\beta_\mathrm{XRT}=1.0\pm 0.3$ we get $\alpha=3\,\beta/2=1.5\pm 0.45$
(c.f. $\alpha=1.68\pm 0.02$ in Table~\ref{tab:100906_breaks}, the 1st
row, before the break). The electron spectral index is $p=2\,\beta+1 =
3.0\pm 0.6$. After the break, applying a uniform spreading jet
scenario,we arrive at $\alpha=2\,\beta+1=3.2\pm 0.4$. The electron
spectral index has the same functional form $\beta=(p-1)/2$. The value
of $\beta$ is almost the same, ($\beta_\mathrm{XRT}=1.1\pm 0.2$), and we
arrive at $p=3.2\pm 0.4$. Thus the closure relations can be satisfied
for a consistent value of $p$. The relations applied imply that
$\nu_\mathrm{m}<\nu_\mathrm{x}< \nu_\mathrm{c}$, where $\nu_\mathrm{m}$
and $\nu_\mathrm{c}$ are the characteristic synchrotron and cooling
frequencies, respectively.

We have tried to find out whether this break is achromatic. The OSN data $R$
show that around 52~ks there is an abruptness of the $R$ light curve
(the fit parameters in the row 2 of Table~\ref{tab:100906_breaks}).
But the time sampling of OSN $R$ data alone is not good enough to clear up
the behavior around this time. If we combine the OSN and NOT data (the
3rd row in Table~\ref{tab:100906_breaks}), we 
apparently do not get an achromatic peak as the tentative break on the $R$
light curve shifts to about 35~ks. 

\begin{figure}
\includegraphics[width=0.45\textwidth]{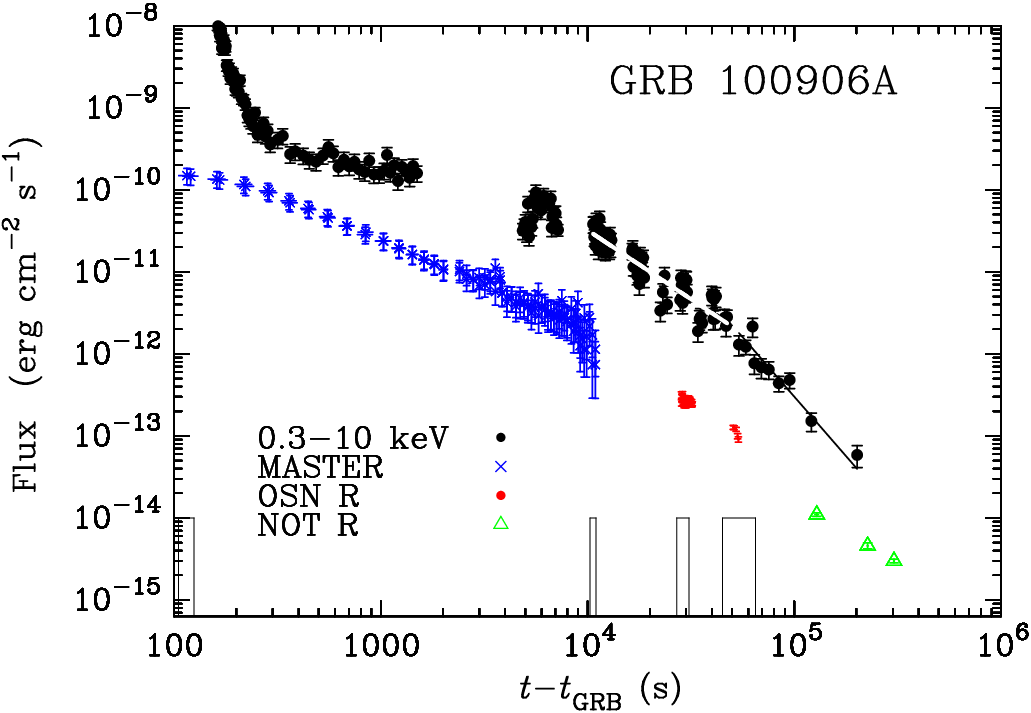}
\caption{GRB\,100906A light curves: unabsorbed 0.3-10~keV flux (black
dots) observed by {\em Swift}/XRT, unfiltered MASTER band flux (blue
crosses), $R$-band flux by the OSN (red dots), and R-band flux by the
NOT (green triangles). The optical points are corrected for the Galactic
extinction. Broken-power law fit is plotted over the 0.3--10~keV light
curve in white and black color, before and after the break time at
51~ks, respectively. Thin-line rectangles show the time intervals
selected for spectral analysis. }
\label{fig:break_time}
\end{figure}

The slope of the $R$ light curve (OSN+NOT) before the break is obtained
by the data, crowded around 30~ks, and might be misleading. If we
suggest that the slope of the $R$ light curve before the break equals to
the slope of the MASTER light curve for the earlier data (blue crosses
in Fig.~\ref{fig:break_time}), that is $1.08$ (also reported by
\citealt{gcn11235}), the break time shifts to $39.0_{-1}^{+0.5}$~ks (the
temporal slope after the break remains the same $\sim 2.2$ as in the
4th row of Table~\ref{tab:100906_breaks}.).

We conclude that the X-ray data alone indicates a jet break that can 
satisfy the closure relations. The optical data do not yield
confidence about the achromatic character of the break. For the
discussion of Amati and Ghirlanda relations, see
\S~\ref{s:amati_gir_06}.

\section{Discussion}
\label{section:discussion}

\subsection{Optical emission sites of GRBs\,100901A and 100906A}
\label{section:opt_sites}

The prompt behavior of the two closely studied GRBs is quite distinct. 
 For GRB\,100901A, we see a similarity of
light curves in the optical, $\gamma$, and X-rays at around 300--500~s
after the trigger. For GRB\,100906A, such similarity of temporal
behaviour is  absent. It has a perfectly smooth optical light
curve, gradually rising and decaying, a different profile from that in
the 15--150~keV band~(Fig.~\ref{fig:lc100906a_all}).

From this point of view, GRBs 100901A and 100906A support the earlier
noticed diversification for optical emission detected during the first
few minutes after the onset of a
GRB~\citep[e.g.,][]{vestrand_etal2005,vestrand_etal2006}: in some cases
the prompt optical flux variations are correlated with the prompt
$\gamma$-emission, and in other cases early optical afterglow light
curve is uncorrelated with the $\gamma$-ray light curve.

Such division of GRBs can be understood within the picture of the
optical emission generated from different sites of the relativistic jet:
in the internal, external forward, or reverse shock
waves~\citep[e.g.,][]{sari-piran1999,meszaros-rees1999}. The external
forward shocks arise due to the interaction of the ejected material with
the ISM and propagate into the ambient media, the reverse shock
propagates into the ejecta itself, and the internal shocks are produced
because of the collisions among many engine-ejected
shells~\citep[e.g.,][]{meszaros-rees1997,sari-piran1997}. The
correlation between the optical, $\gamma$, and X-rays can be achieved,
if the optical emission is produced by the synchrotron mechanism within
the same region as the hard radiation.

The early X-ray and soft $\gamma$-ray spectrum of GRB\,100901A shows a
dependence close to $ \nu ^{-1/2} $ which is typical for the synchrotron
radiation if $\nu_\mathrm{c}< \nu < \nu_\mathrm{m}$ (where
$\nu_\mathrm{m}$ and $\nu_\mathrm{c}$ are the characteristic synchrotron
and cooling frequencies, respectively) in the regime when electrons lose
energy very quickly~\citep[see, e.g.,][]{sari_etal1998,sari-piran1999}.
Adding the MASTER optical point (corrected for the Galaxy extinction),
and fitting simultaneously the optical, $0.3-10$~keV, and $15-150$~keV
data by an absorbed power law, we can determine the magnitude of the
optical extinction in the host. Consistent values of $A_V^\mathrm{int}$
and $N_\mathrm{H}^\mathrm{int}$ are obtained for the prompt spectral
distributions (Table~\ref{tab:spectral_indexes}), indicating a
persisting relation between the optical and high energy part of the
spectrum. Basically, this means that the optical emission is the
low-energetic photons from the same region that generates the high
energy photons.

We should add that the data can also be reconciled with the prompt
spectral distributions of GRB\,100901A peaking between the optical and
X-rays, if $\nu_\mathrm{opt}<\nu_\mathrm{c}<\nu_\mathrm{x}$. Then, a 
migration of the cooling frequency $\nu_\mathrm{c}$ ($\propto t^{-1/2}$,
see, for example, \citealt{sari_etal1998}) may be hidden in the errors
of resulted $A_V^{\mathrm{int}}$. Spectral fitting of an absorbed power
law to the late-time MASTER and $0.3-10$~keV data provides
$A_V^\mathrm{int}=0.45\pm 0.15$, lower than $A_V^\mathrm{int}=0.8\pm
0.1$ for the prompt fits (values are for the SMC type of the extinction
law; see \S~\ref{section:ext01}). This could be a confirmation of the
above suggestion, but a detailed discussion is beyond the aims of the
present study.

In contrast to the above picture, GRB\,100906A exhibits no correlation
between the optical and high energy emission at 15--150~keV. We suggest
that in this case the optical radiation originates at the front of the
external shock, and the light curve shows a smooth conversion of a bow
shock to a self-similar mode of the afterglow with a power decay
$F_\mathrm{opt} \propto t ^{-\alpha}$ with $\alpha=
1.08\pm0.01$~(Table~\ref{tab:100906_breaks}, see also
\citealt{gcn11235}). 

A critical test for determining the site of the optical emission could
be the observation of polarization. If radiation in the internal shock
occurs in the ordered magnetic field, some polarization is
expected~\citep[see, e.g.,][]{steele_etal2009}. This implies that the
magnetic field coherence scale is larger than the size of the visible
emitting region. Thus, a discovery of a considerable ($> 10\%$)
polarization of the optical radiation, correlating with the prompt
$\gamma$-ray emission, would be a strong evidence in favor of the
optical radiation originating in the internal shock. Conversely, the
afterglow radiation from the external shock occurs in the chaotic field
of the interstellar medium compressed by the shock and should not have a
considerable polarization~\citep{gruzinov-waxman1999}. 
Low degree of polarization of the optical light would rule
out the presence of a large-scale ordered magnetic field in the emitting
region~\citep{mundell_etal2007}.

Fig.~\ref{fig:lc100906a_opt} shows our simultaneous observations of
GRB\,100906A with two orthogonal polaroids. The signals in the two
channels are the same with an accuracy of 0.5 per cent. Absence of the
strong polarization is consistent with the picture of the different
producing sites of the high-energy and optical radiation in
GRB\,100906A. However, we should note that the MASTER polarization
observations of GRB\,100906A were made with two optical tubes of one
telescope. Such mode of observations is restricted in the sence of
giving definite information on the linear polarization (see the
Appendix). Polarization filters are oriented differently in the
celestial coordinate system at the telescopes of the MASTER net. A
future observation of a suitable GRB by two MASTER telescopes should
thus provide an accurate polarization information. For the observations
of GRB\,100906A, we can infer that in the time interval from 100~s to
$\sim 1$~h the degree of linear polarization of the optical emission is
less than $10\%$ with probability of about 60 per cent (see the Appendix
and Fig.~\ref{fig:prob_polarization}). The full information of the
inferred fractional linear polarization can be retrieved from
Fig.~\ref{fig:prob_polarization}. We note that the probabilities are
obtained under a very conservative assumption that a source can have any
fractional linear polarization from 0 to 100\%.

\begin{figure*}
\includegraphics[width=150mm]{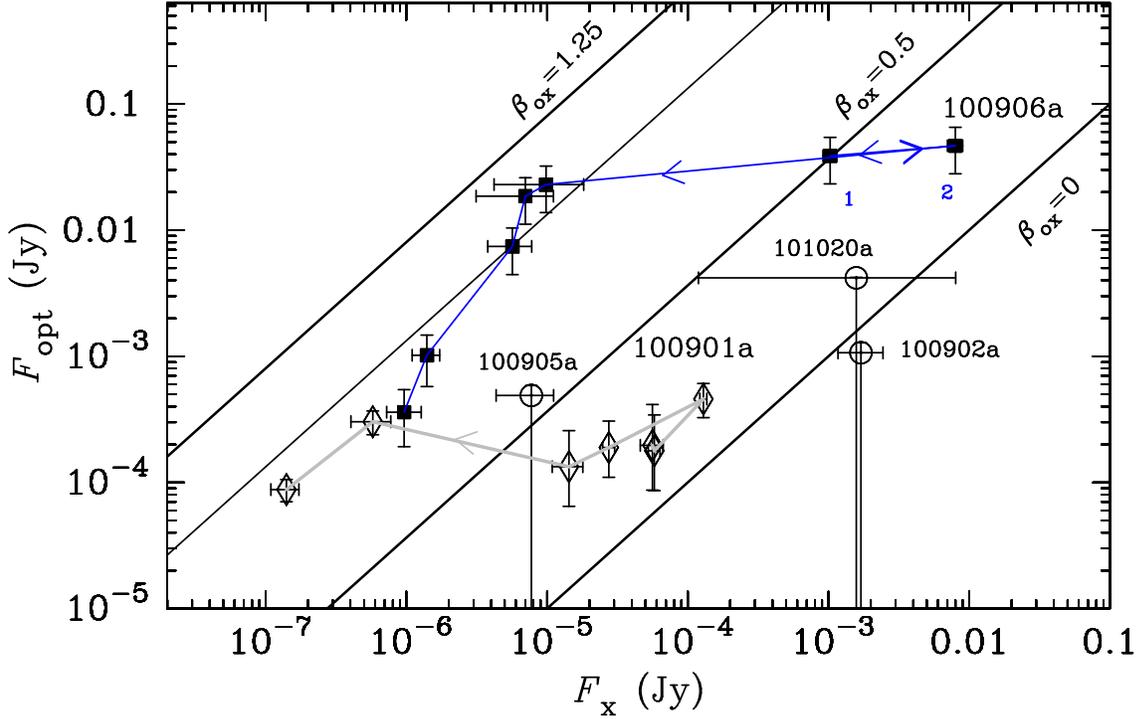} 
\caption{Diagram of the
optical flux density versus X-ray flux density at 3 keV for five GRBs observed by
MASTER
in September and October 2010. Each point represents a
simultaneous observation in the optical and by {\em Swift}/XRT. The optical
flux densities have been corrected for the Galactic extinction. 
Diamonds
connected with grey line show evolution of GRB\,100901A (eight time intervals:
193--233,
 312--372,
 387--467,
 481--581,
 595--715,
 3800--4150,
 27000--28500, and
 102800--121500 s); squares
connected by blue lines, of GRB\,100906A (seven time intervals: 
84--94 (point 1),
105--125 (point 2),
330--390,
410--480, 
942--1112,
6000--7000, and 
10300--11000~s).
 For GRBs 100902A,
100905A, and 101020A, only upper limits on optical flux density are given at
post-trigger times 132--162, 150--180, and 106--126~s, respectively. 
When tracks
are parallel to $\beta=$const lines, the optical flux density is emphatically
correlated with the 3~keV flux density. We adopt Galactic extinctions 
$A_V =1.037$ for GRB\,100902A,
$A_V =0.191$ for GRB\,100905A, and $A_V =0.051$ for GRB\,101020A from
the NED Extragalactic Calculator. 
\label{fig:beta_all}}
\end{figure*}

\subsection{General spectral properties}
\label{subsection:sp_evol}

In order to view spectral characteristics of the observed GRBs at one
glance, we construct in Fig.~\ref{fig:beta_all} a diagram of the optical
flux density versus X-ray flux density at 3~keV, following
\citet{Jakobsson_etal2004}. However, we plot not the values extrapolated
to 11~h time mark, but rather the evolution for each GRB. Here, a
spectral index $\beta$ ($F_{\nu} \propto \nu^{-\beta}$) is shown by
straight lines. Each optical observation is corrected for the Galactic
extinction at a specific waveband using the NED Galactic Extinction
calculator. As the intrinsic spectrum is not known a-priori, we do not
include a correction for the host extinction. The 3~keV flux densities
are the best-fitting values obtained with an absorbed power law model in
the XRT/{\em Swift} energy band. For GRBs\,100902A, 100905A, and
101020A, only upper optical limits are shown.

The limit for GRB\,100902A at $132-162$~s yields that the burst was
rather X-ray bright and very faint in the optical waveband at the time,
possibly due to a large intrinsic extinction. Provided that $T_{90} =
429$~s~\citep{gcn11202}, one can expect that the GRB would have low
$\beta_\mathrm{ox}$ at $(0.3-0.4)\,T_{90}$. It might be a `dark' GRB
according to classification by \citet{Jakobsson_etal2004}, bearing in
mind its very prominent position on the diagram. GRB\,100902A
appears to be darker than any of the GRBs of the sample of \citet[][see
their Fig.~1]{Jakobsson_etal2004}, with all of them being above the
$\beta_\mathrm{ox} = 0$ line.

 An anti-correlation between excess absorption column densities and
redshifts of GRBs is expected due to the shift of the source rest-frame
energy range below 1~keV out of the X-Ray Telescope observable energy
range. Thus \citet{grupe_etal2007} suggest a relation to estimate an upper
limit on the redshift. From the late-time spectrum of GRB\,100902A, one
finds an excess absorption column density of $\sim 2.8^{+0.6}_{-0.7}
\times 10^{21}$~cm$^{-2}$ and upper limit for $z\approx 3.2$
(substituting lower value of $N_\mathrm{H}$ from expression (1) of
\cite{grupe_etal2007}). High redshifts are thought to be possibly
responsible for darkness of GRBs, but the redshift obtained above is not
high enough for this scenario~\citep[see, e.g.,][and references
therein]{Jakobsson_etal2004}; on the other hand, expression by
\citet{grupe_etal2007} is inferred for non-dark GRBs and might be
unapplicable. We also note that analysis of \cite{gcn11195} who use
a cutoff power-law model suggest a higher redshift, $z\sim 4.5$.

The MASTER observation of GRB\,100905A was done at the time interval
150--180~s that is much later than its $T_{90}=3.4$~s. 
It can belong to the class of intermediate duration
GRBs~\citep{horvath1998,horvath2002,Mukherjee_etal1998}. At 150--180~s 
after the trigger its XRT spectral slope is $\beta_\mathrm{x}=0.2\pm
0.2$. If the optical emission belong to the same spectral distribution,
it would explain its weakness, because the line $\beta_\mathrm{ox}=0.2$
intersects the region of the 100905A on the diagram at the
$F_\mathrm{opt}$ values well below the optical limit.

GRB\,101020A could not be observed by {\em Swift}/XRT. The flux density
at 3~keV for time $0.65\times T_{90}$, where $T_{90}=177$~s, is
extrapolated from the BAT data and has a considerable uncertainty,  thus
resulting in a wide area in Fig.~\ref{fig:beta_all}. Nevertheless, the
optical-to-X-ray flux ratio can be similar to that of GRBs\,100901A and
100906A.

We see from Fig.~\ref{fig:beta_all} that the flux density ratios of
GRB\,100901A and GRB\,100906A ultimately tend to the area between
$\beta=0.5$ and 1.25. This is consistent with the fireball model
\citep[e.g.][]{sari_etal1998}. Spectral slope $\sim -1$ at late times is
consistent with the prediction for the slow-cooling regime for
frequencies higher than the cooling frequency for electrons with
distribution of Lorentz factors $N(\gamma_e) \propto \gamma_e^{-p}$ with
$p\ga 2$. 

The distinct prompt behavior of GRB\,100901A and GRB\,100906A is evident
in the diagram. We witness the optical and X-ray flux density
correlation in GRB\,100901A at the time of the last $\gamma$-ray pulse, 
together with an approximately constant spectral index. In GRB\,100906A,
during the $\gamma$-ray pulse, occurring about 100~s after the trigger, 
the 3~keV flux density undergoes a strong variation. The same can be
said about the spectral indices in the XRT and BAT bands. However, the
optical flux density is roughly at the same level. We remind that the
spectral fitting provides an evidence that the prompt high energy SED
has a peak, which migrates from about 30 to 4~keV at times 80-130~s
after $T0$ (see \S~\ref{s.lc_spectra}). It should be added that during
those times the X-ray emission in the interval of 0.3-10~keV can be
contributed by the two different sites with a proportion varying with
time: the one that is responsible for the $\gamma$-rays and one
producing the optical photons.

\subsection{Amati and Ghirlanda relations for GRB\,100901A}

The total duration BAT spectrum of GRB\,100901A yields $E_\mathrm{iso} =
6.3 \times 10^{52}$~erg in the energy interval $1-10^4$~keV in the rest
frame. From the Amati relation~\citep{amati2006} one gets the expected
position of of the energy distribution peak in the rest frame
$E_\mathrm{p,i} = 230$~keV, which translated to the observer frame
becomes $E_\mathrm{p}\sim 100$~keV. This does not contradict with the
lower limit $E_\mathrm{p}^\mathrm{min} \sim 75$ keV, derived on the
basis that no characteristic energy can be found form the spectral
analysis of the BAT data, and the photon index of the power law is about
-1.5.

The 0.3--10~keV light curve 
$\propto t^{-1.5}$ from $4\times 10^{4}$~s to about
$10^{6}$~s when a possible steepening occurs. The MASTER $W$-band light
curve enables only a rough evaluation of 
temporal index in the interval $3\times 10^{4} - 3\times 10^{5}$~s: it
is decaying approximately as $\sim t^{-1.3}$ with just one late optical
point at $\sim 10^5$~s.
Using the tentative jet break time $\sim 10^6$~s, one gets the opening
half-angle of the jet $\sim 13$~deg for 
efficiency of converting the ejecta energy into $\gamma$-rays
$\eta_\gamma=0.2$ and the number density of the ambient medium
$n=3$~cm$^2$. Few jets are reported to have comparably wide
angles~\citep{Ghirlanda_etal2004}.
Corresponding  collimation-corrected energy would be $E_\gamma =
1.7\times 10^{51}$~erg, which using Ghirlanda relation predicts
$E_\mathrm{p,i} \sim 700$~keV, a value about 3 times higher than prediction of
the Amati relation.

We conclude that GRB\,100901 can be consistent with the Amati relation,
if $E_\mathrm{p,i} \sim 230$~keV, but the Ghirlanda relation is hardly
applicable.

\subsection{Amati and Ghirlanda relations for GRB\,100906A}
\label{s:amati_gir_06}

 \citet{gcn11251}, fitting the spectrum of GRB\,100906A in the time interval
$3-142$~s with the Band function, derive values
$E_\mathrm{peak} =142_{-60}^{+119}$~keV and $E_\mathrm{iso}=(2.2\pm 0.4)
\times 10^{53}$~erg.
The peak of the energy distribution  in the
rest frame is at $E_\mathrm{p,i}= (1+z)\times E_\mathrm{peak} =
387_{-164}^{+324}$~keV. Central values of
$E_\mathrm{iso}$ and $E_\mathrm{p,i}$ are  close to the Amati
correlation $E_\mathrm{p,i} = 95 \times E^{0.49}_{\mathrm{iso}}$
(figure 2 of \citealt{amati2006}).

 If we assume  jet break time $t_\mathrm{b}=51\pm4$~ks (see
\S~\ref{section:jetbreak}) and
substitute the above $E_\mathrm{iso}$ into equation (1)
of~\citet{Ghirlanda_etal2004} (where we assume the  values of the
efficiency of converting the ejecta energy into $\gamma$-rays
$\eta_\gamma=0.2$ and the number density of the ambient medium
$n=3$~cm$^2$), we obtain the opening half angle of the jet
$\theta_\mathrm{j} =3.31\pm 0.08$~deg. Then the collimation-corrected
energy is $E_\gamma \sim (3.7\pm 0.7) \times 10^{50}$~erg. The last value
together with $E_\mathrm{p,i}$  agree with the Ghirlanda relation
$E_\mathrm{p,i} \simeq 480
(E_\gamma/10^{51}~\mathrm{erg})^{0.7}$~(figure 1 of
\citealt{Ghirlanda_etal2004}).  This is especially so due to the big uncertainty in the
observed $E_\mathrm{peak}$ value of GRB\,100906A.

\subsection{Long life of a central engine}

A prolonged activity of the
central engine was suggested as a source  of an extended afterglow by
\citet{katz-piran1997}.
 It was considered later as an explanation for
plateaus on X-ray light curves of some
GRBs~\citep[e.g.,][]{zhang-meszaros2001,zhang_etal2006,liang_etal2007,
willingale_etal2007,lipunov-gorb2008}. In particular, sharp drops at
late times on the light curves indicate that such variations are likely
due to some internal mechanisms~\citep{troja_etal2007,Zhang2009}.
To account for a lasting activity of the central engine of GRBs, a model
of a rapidly-spinning, highly-magnetized neutron star, a
`proto-magnetar', formed in the result of a core collapse, was
proposed~\citep[see review by][]{Metzger2010}. Another object that can
provide a long-term energy supply is a rapidly-spinning, magnetized
collapsed core with mass exceeding an upper limit for a neutron star.
Its fast rotation stops the collapse, and a quasi-stationary
configuration is presumably formed at this stage, which is called a
`spinar'~\citep{lipunova1997e,lipunov-gorbovskoy2007}. Before a collapse
to a black hole is possible, enough angular momentum should be taken out
from the system. The time required for angular momentum dissipation
determines the duration of the central engine activity and depends on
the magnetic field strength, as the energy and angular momentum losses
are governed by the magneto-dipole mechanism.

\begin{figure}
\includegraphics[width=84mm]{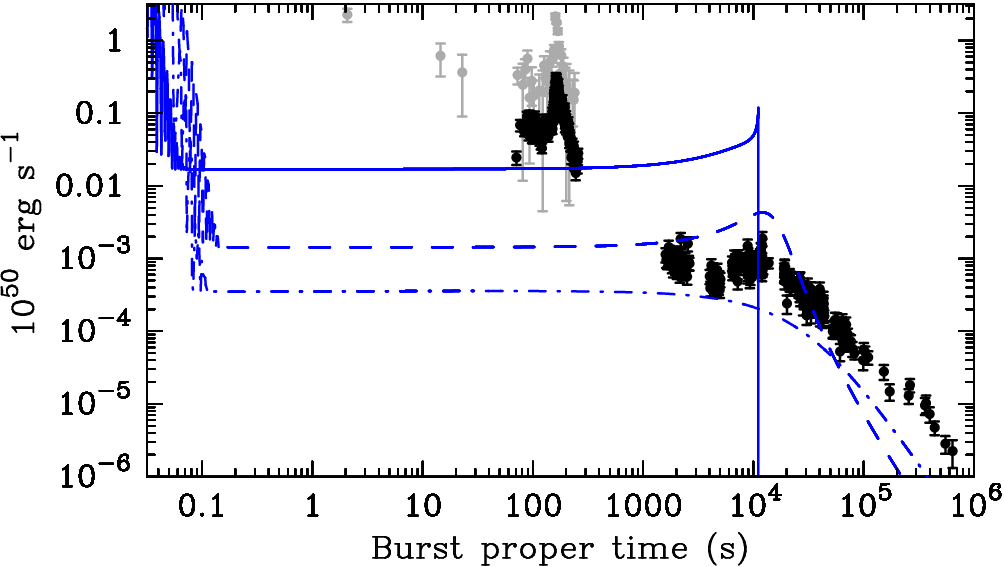} 
\caption{ Energy release
of a collapsing core calculated following \citet{lipunov-gorb2008}.
Points with bars are the observed 0.3--10~keV and 15--150~keV flux from
GRB\,100901A \citep{gcn11169b,gcn11171} translated to redshift $1.408$.
Three curves with plateaus represent numerical solutions for energy
release rates, multiplied by 0.1, for cores of different mass and
angular momentum. The dashed and dot-dashed lines are solutions for
proto-magnetars of the same mass and with different angular momentum at
the halt of the initial dynamical collapse that produced the first peak
before the plateau. The first peak is referred to time zero; the details
of the first peak are of no relevance. The top solid line shows a
solution for a collapsing core with such a mass that makes a collapse to
a black hole inevitable. For specific parameters of the models, see
Discussion.
\label{fig:100901a_model}}
\end{figure}

 \citet{lipunov-gorb2008} proposed a set of \hbox{1-D} equations
determining evolution of a collapsing rotating magnetized core with main
relativistic effects taken into account. These equations describe the
principal behavior of both the proto-magnetar and the spinar. Solutions
suggest two peaks in energy release: the first takes place when
dynamical collapse is halted by centrifugal or pressure forces, and the
second, when final contraction occurs. The peaks are separated by a
plateau. Three exemplary solutions of equations by
\citet{lipunov-gorb2008} are shown in Fig.~\ref{fig:100901a_model}. The
first peak is set to occur at zero time and not shown in detail. In the
model, the first peak is evaluated on a dynamical time-scale and does
not involve specific jet and afterglow radiation mechanisms. If the
first stage of a collapse finishes close to $r_\mathrm{g}=G\,M/c^2$,
then the second peak is extinguished. A power of a proto-magnetar,
defined by the pulsar dipole mechanism, approaches the $t^{-2}$ law
after the second peak. A power of a spinar has always a very sharp drop
after the second peak. For each of the models in
Fig.~\ref{fig:100901a_model}, calculated power is multiplied by an
arbitrary factor of 0.1.

Two curves with lower plateaus represent models with mass
$M=1.9~M_\odot$. They both go as $\propto t^{-2}$ at late times. In the
model with monotonic transition to a decay, a proto-magnetar forms with
period $P=2$~ms and poloidal magnetic field $B \sim 4\times 10^{14}$~G
at $\sim 13 \,r_\mathrm{g}$. The middle plateau represents a newly born
magnetar of higher angular momentum: the dynamical collapse halts at
about $40\,r_\mathrm{g}$ with $P=5$~ms and $B\sim 7\times 10^{13}$~G.
The object contracts further, to $10\,r_\mathrm{g}$, causing a spin up
to $P=1.7$~ms and the dipole magnetic field amplification up to $\sim
10^{15}$~G. This specific evolution manifests itself as a bump at the
end of the plateau.
The solid curve shows a solution for a spinar with $M=4~M_\odot$. At the
plateau, the dimensionless Kerr parameter $a_\mathrm{Kerr}=Jc/GM^2=3$,
$P\sim 1$~ms, $B\sim 1.5\times 10^{13}$~G, and the characteristic radius
$15 \,r_\mathrm{g}$. The spike at the end is numerically solved and
indicates the collapse to a black hole. The initial magnetic field
strengths are chosen in order to obtain the characteristic time of a
plateau $\sim 10^4$~s.

A spinar and a proto-magnetar, both can explain long activity of GRBs. 
The law of temporal decay after the second peak (or the end of a
plateau) can be defined by an afterglow and does not allow us to decide
which compact object is behind the scene. The efficiency factor, which
is set 0.1 for Fig.\ref{fig:100901a_model}, cannot be constrained in our
general approach.

\section{Summary}
\label{section:summary}

The net of optical robotic telescopes MASTER coming into full operation
over the Russia territory is encouraged by the need of dense early
observations of GRBs around the globe. Optical ground-based
observations, with polarization measurements, are indispensable in
revealing the nature of the GRB central engine. The MASTER team is
working on obtaining optical linear polarization information at the
earliest time since a GRB trigger, before $\gamma$-emission ceases, by
observing simultaneously with several telescopes. Our goal is to deliver
full polarization information for GRBs in the optical band as early as
possible after the trigger time.

Here we presented the early results of the MASTER network. During 2010,
MASTER observed five GRBs described in this paper. Optical transients
were found for two bursts: GRBs\,100901A and GRB\,100906A. Their
detailed analysis was done basing on the MASTER data, the data from the
1.5-m telescope at Sierra Nevada Observatory and 2.56-m Nordic Optical
Telescope, as well as the publicly available data from the {\em Swift}
mission.

The light curves and spectra of GRBs\,100901A and 100906A suggest
different origin of the prompt optical emission. The correlation between
the optical and high-energy light curves of GRB\,100901A, as well as the
results of spectral fitting favor the hypothesis of the common origin
for the optical and high-energy spectrum. During the $\gamma$-ray flare
started at about 300~s after the trigger, the slope of the spectral
distribution from the optical to 150~keV is close to $-1/2$. This is
expected for the synchrotron radiation at the frequencies above the
cooling and below the characteristic synchrotron frequency in the 'fast
cooling' regime \citep[see, e.g.][]{sari-piran1999}. The resulted values
of the host optical extinction are consistent for the prompt fits and
similar to the values obtained by fitting an absorbed power law to the
late-time spectral data. This means that the prompt optical points lie
on the same spectral distribution which describes the $\gamma$- and
X-ray data, supporting the hypothesis of their common production site
and mechanism. The estimated ratio of the GRB host hydrogen column
density $N_\mathrm{H}^\mathrm{int}$ to the total extinction
$A_V^\mathrm{int}$ suggests that the host of GRB\,100901A possesses the
metal-to-dust ratio comparable to the level measured for the Milky Way
or Magellanic Clouds.

In GRB\,100906A, no apparent correlation between the optical and 
$15-150$~keV light curves is seen. We suggest that in GRB\,100906A the
prompt optical radiation originates in a region different from the
$\gamma$-ray production site, probably, in a front shock. An analysis of
GRB\,100906A reveals that a spectral component, responsible for the
$\gamma$-ray pulse at $\sim 100$~s after the trigger and the downward
migration of the X-ray peak on the spectral energy distribution, does
not correlate with the optical emission at these times. 

Signals in two orthogonal polarizations, measured by the MASTER 
telescope at Tunka from GRB\,100906A, are found to be equal within 0.5
per cent. We calculate the probability distribution for the fractional
linear optical polarization of GRB\,100906A. We point out that a future
observation of a GRB by at least two MASTER telescopes should yield an
accurate polarization information.

The 0.3--10~keV light curve of GRB\,100906A has a break at around 14 hr
after the trigger. The X-ray spectral index (non-variable around the
break) and the temporal slopes can be reconciled with the closure
relations for the relativistic jet in the `slow cooling' regime before
the break and the spreading uniform jet after the break. It is not clear
from the data in the $R$ band if this break is achromatic. Estimated jet
opening angle, $3.31\pm 0.08$~degrees, and collimation-corrected energy,
$E_\gamma \sim (3.7\pm 0.7) \times 10^{50}$~erg, imply that GRB\,100906A
follows the Amati relation and does not contradict the Ghirlanda
relation. An analysis of the late-time spectra provides the upper limit
on the optical extinction in the GRB host.

\section*{Acknowledgments}
 We thank the referee for many useful comments. We are grateful to T.
Piran for his help. We also thank K. Sokolovsky. We acknowledge Taka
Sakamoto and Scott D. Barthelmy for the BAT data.This work was supported
by the Ministry of Science of the Russian Federation (state contract no.
02.740.11.0249). E.S.G. thanks the Foundation for Non-Profit Programs
`Dynasty'. The research  of J.G., A.J.C.T., J.C.T., M.J., and R.S.R. is
supported   by the Spanish programs   AYA2007-63677, AYA2008-03467/ESP
 and AYA2009-14000-C03-01. Authors are grateful to S.M. Bodrov for the
ceaseless support of the MASTER-Net. This work made use of data supplied
by the UK {\em Swift} Science Data Centre at the University of
Leicester. This research has made use of the NASA/IPAC Extragalactic
Database (NED) which is operated by the Jet Propulsion Laboratory,
California Institute of Technology, under contract with the National
Aeronautics and Space Administration. We used the publicly available
results of the Sloan Digital Sky Survey, whose web site is
http://www.sdss.org/.
This research has made use of NASA's Astrophysics Data System.

\bibliographystyle{mn2e}

\begin{thebibliography}{83}
\expandafter\ifx\csname natexlab\endcsname\relax\def\natexlab#1{#1}\fi

\bibitem[{{Abazajian} {et~al}\mbox{.}(2009){Abazajian}, {Adelman-McCarthy},
  {Agueros}, {Allam}, {Allende Prieto}, {An}, {Anderson}, {Anderson}, {Annis},
  {Bahcall}, {Bailer-Jones}, \& {et al}}]{abazajian_etal2009}
{Abazajian} K.~N. {et~al.}, 2009, \apjs, 182, 543

\bibitem[{{Akerlof} {et~al}\mbox{.}(2000){Akerlof}, {Balsano}, {Barthelmy},
  {Bloch}, {Butterworth}, {Casperson}, {Cline}, {Fletcher}, \& {et
  al}}]{akerlof_etal2000}
{Akerlof} C. {et~al.}, 2000, \apjl, 532, L25

\bibitem[{{Amati}(2006)}]{amati2006}
{Amati} L., 2006, \mnras, 372, 233

\bibitem[{{Arnaud}(1996)}]{arnaud1996}
{Arnaud} K.~A., 1996, in Astronomical Society of the Pacific Conference Series,
  Vol. 101, Astronomical Data Analysis Software and Systems V, {G.~H.~Jacoby \&
  J.~Barnes}, ed., pp. 17--20

\bibitem[{{Atwood} {et~al}\mbox{.}(2009){Atwood}, {Abdo}, {Ackermann},
  {Althouse}, {Anderson}, {Axelsson}, {Baldini}, {Ballet}, {Band},
  {Barbiellini}, {Bartelt}, {Bastieri}, {Baughman}, {Bechtol},
  {B{\'e}d{\'e}r{\`e}de}, {Bellardi}, {Bellazzini}, {Berenji}, {Bignami},
  {Bisello}, {Bissaldi}, {Blandford}, {Bloom}, {Bogart}, {Bonamente},
  {Bonnell}, {Borgland}, {Bouvier}, {Bregeon}, {Brez}, {Brigida}, {Bruel},
  {Burnett}, {Busetto}, {Caliandro}, {Cameron}, {Caraveo}, {Carius}, {Carlson},
  {Casandjian}, {Cavazzuti}, {Ceccanti}, {Cecchi}, {Charles}, {Chekhtman},
  {Cheung}, {Chiang}, {Chipaux}, {Cillis}, {Ciprini}, {Claus}, {Cohen-Tanugi},
  {Condamoor}, {Conrad}, {Corbet}, {Corucci}, {Costamante}, {Cutini}, {Davis},
  {Decotigny}, {DeKlotz}, {Dermer}, {de Angelis}, {Digel}, {do Couto e Silva},
  {Drell}, {Dubois}, {Dumora}, {Edmonds}, {Fabiani}, {Farnier}, {Favuzzi},
  {Flath}, {Fleury}, {Focke}, {Funk}, {Fusco}, {Gargano}, {Gasparrini},
  {Gehrels}, {Gentit}, {Germani}, {Giebels}, {Giglietto}, {Giommi}, {Giordano},
  {Glanzman}, {Godfrey}, {Grenier}, {Grondin}, {Grove}, {Guillemot}, {Guiriec},
  {Haller}, {Harding}, {Hart}, {Hays}, {Healey}, {Hirayama}, {Hjalmarsdotter},
  {Horn}, {Hughes}, {J{\'o}hannesson}, {Johansson}, {Johnson}, {Johnson},
  {Johnson}, {Johnson}, {Kamae}, {Katagiri}, {Kataoka}, {Kavelaars}, {Kawai},
  {Kelly}, {Kerr}, {Klamra}, {Kn{\"o}dlseder}, {Kocian}, {Komin}, {Kuehn},
  {Kuss}, {Landriu}, {Latronico}, {Lee}, {Lee}, {Lemoine-Goumard}, {Lionetto},
  {Longo}, {Loparco}, {Lott}, {Lovellette}, {Lubrano}, {Madejski}, {Makeev},
  {Marangelli}, {Massai}, {Mazziotta}, {McEnery}, {Menon}, {Meurer},
  {Michelson}, {Minuti}, {Mirizzi}, {Mitthumsiri}, {Mizuno}, {Moiseev},
  {Monte}, {Monzani}, {Moretti}, {Morselli}, {Moskalenko}, {Murgia},
  {Nakamori}, {Nishino}, {Nolan}, {Norris}, {Nuss}, {Ohno}, {Ohsugi}, {Omodei},
  {Orlando}, {Ormes}, {Paccagnella}, {Paneque}, {Panetta}, {Parent}, {Pearce},
  {Pepe}, {Perazzo}, {Pesce-Rollins}, {Picozza}, {Pieri}, {Pinchera}, {Piron},
  {Porter}, {Poupard}, {Rain{\`o}}, {Rando}, {Rapposelli}, {Razzano}, {Reimer},
  {Reimer}, {Reposeur}, {Reyes}, {Ritz}, {Rochester}, {Rodriguez}, {Romani},
  {Roth}, {Russell}, {Ryde}, {Sabatini}, {Sadrozinski}, {Sanchez}, {Sander},
  {Sapozhnikov}, {Parkinson}, {Scargle}, {Schalk}, {Scolieri}, {Sgr{\`o}},
  {Share}, {Shaw}, {Shimokawabe}, {Shrader}, {Sierpowska-Bartosik}, {Siskind},
  {Smith}, {Smith}, {Spandre}, {Spinelli}, {Starck}, {Stephens}, {Strickman},
  {Strong}, {Suson}, {Tajima}, {Takahashi}, {Takahashi}, {Tanaka}, {Tenze},
  {Tether}, {Thayer}, {Thayer}, {Thompson}, {Tibaldo}, {Tibolla}, {Torres},
  {Tosti}, {Tramacere}, {Turri}, {Usher}, {Vilchez}, {Vitale}, {Wang},
  {Watters}, {Winer}, {Wood}, {Ylinen}, \& {Ziegler}}]{Fermi2009}
{Atwood} W.~B. {et~al.}, 2009, \apj, 697, 1071

\bibitem[{{Barthelmy} {et~al}\mbox{.}(2010{\natexlab{a}}){Barthelmy},
  {Baumgartner}, {Cummings}, {Gehrels}, {Krimm}, {Markwardt}, {Marshall},
  {Palmer}, {Sakamoto}, {Stamatikos}, {Tueller}, \& {Ukwatta}}]{gcn11218}
{Barthelmy} S.~D. {et~al.}, 2010{\natexlab{a}}, GRB Coordinates Network,
  Circular Service, 11218, 1 (2010), 1218, 1

\bibitem[{{Barthelmy} {et~al}\mbox{.}(2010{\natexlab{b}}){Barthelmy},
  {Baumgartner}, {Cummings}, {Gehrels}, {Krimm}, {Markwardt}, {Palmer},
  {Sakamoto}, {Stamatikos}, {Tueller}, \& {Ukwatta}}]{gcn11233}
---, 2010{\natexlab{b}}, GRB Coordinates Network, Circular Service, 11233, 1
  (2010), 1233, 1

\bibitem[{{Beardmore} \& {Markwardt}(2010)}]{gcn11244}
{Beardmore} A.~P., {Markwardt} C.~B., 2010, GRB Coordinates Network, Circular
  Service, 11244, 1 (2010), 1244, 1

\bibitem[{{Butler} \& {Kocevski}(2007)}]{Butler-Kocevski2007}
{Butler} N.~R., {Kocevski} D., 2007, \apj, 663, 407

\bibitem[{{Campana} {et~al}\mbox{.}(2010){Campana}, {Starling}, {Evans}, \&
  {Sakamoto}}]{gcn11195}
{Campana} S., {Starling} R.~L.~C., {Evans} P.~A., {Sakamoto} T., 2010, GRB
  Coordinates Network, Circular Service, 11195, 1 (2010), 1195, 1

\bibitem[{{Chornock} {et~al}\mbox{.}(2010){Chornock}, {Berger}, {Fox}, {Levan},
  {Tanvir}, \& {Wiersema}}]{gcn11164}
{Chornock} R., {Berger} E., {Fox} D., {Levan} A.~J., {Tanvir} N.~R., {Wiersema}
  K., 2010, GRB Coordinates Network, Circular Service, 11164, 1 (2010), 1164, 1

\bibitem[{{de Ugarte Postigo} {et~al}\mbox{.}(2011){de Ugarte Postigo},
  {Horv{\'a}th}, {Veres}, {Bagoly}, {Kann}, {Th{\"o}ne}, {Balazs}, {D'Avanzo},
  {Aloy}, {Foley}, {Campana}, {Mao}, {Jakobsson}, {Covino}, {Fynbo},
  {Gorosabel}, {Castro-Tirado}, {Amati}, \& {Nardini}}]{postigo_etal2011}
{de Ugarte Postigo} A. {et~al.}, 2011, \aap, 525, A109

\bibitem[{{Evans} {et~al}\mbox{.}(2009){Evans}, {Beardmore}, {Page}, {Osborne},
  {O'Brien}, {Willingale}, {Starling}, {Burrows}, {Godet}, \& {et
  al}}]{evans_etal2009}
{Evans} P.~A. {et~al.}, 2009, \mnras, 397, 1177

\bibitem[{{Evans} {et~al}\mbox{.}(2010){Evans}, {Willingale}, {Osborne},
  {O'Brien}, {Page}, {Markwardt}, {Barthelmy}, {Beardmore}, {Burrows},
  {Pagani}, {Starling}, {Gehrels}, \& {Romano}}]{evans_etal2010}
---, 2010, \aap, 519, A102

\bibitem[{{Everett} \& {Howell}(2001)}]{everett-howell2001}
{Everett} M.~E., {Howell} S.~B., 2001, \pasp, 113, 1428

\bibitem[{{Gehrels} {et~al}\mbox{.}(2004){Gehrels}, {Chincarini}, {Giommi},
  {Mason}, {Nousek}, {Wells}, {White}, {Barthelmy}, {Burrows}, {Cominsky},
  {Hurley}, {Marshall}, {M{\'e}sz{\'a}ros}, {Roming}, {Angelini}, {Barbier},
  {Belloni}, {Campana}, {Caraveo}, {Chester}, {Citterio}, {Cline}, {Cropper},
  {Cummings}, {Dean}, {Feigelson}, {Fenimore}, {Frail}, {Fruchter}, {Garmire},
  {Gendreau}, {Ghisellini}, {Greiner}, {Hill}, {Hunsberger}, {Krimm},
  {Kulkarni}, {Kumar}, {Lebrun}, {Lloyd-Ronning}, {Markwardt}, {Mattson},
  {Mushotzky}, {Norris}, {Osborne}, {Paczynski}, {Palmer}, {Park}, {Parsons},
  {Paul}, {Rees}, {Reynolds}, {Rhoads}, {Sasseen}, {Schaefer}, {Short},
  {Smale}, {Smith}, {Stella}, {Tagliaferri}, {Takahashi}, {Tashiro},
  {Townsley}, {Tueller}, {Turner}, {Vietri}, {Voges}, {Ward}, {Willingale},
  {Zerbi}, \& {Zhang}}]{Swift2004}
{Gehrels} N. {et~al.}, 2004, \apj, 611, 1005

\bibitem[{{Ghirlanda}, {Ghisellini} \& {Lazzati}(2004){Ghirlanda},
  {Ghisellini}, \& {Lazzati}}]{Ghirlanda_etal2004}
{Ghirlanda} G., {Ghisellini} G., {Lazzati} D., 2004, \apj, 616, 331

\bibitem[{{Golenetskii} {et~al}\mbox{.}(2010){Golenetskii}, {Aptekar},
  {Frederiks}, {Mazets}, {Pal'Shin}, {Oleynik}, {Ulanov}, {Svinkin}, \&
  {Cline}}]{gcn11251}
{Golenetskii} S. {et~al.}, 2010, GRB Coordinates Network, Circular Service,
  11251, 1 (2010), 1251, 1

\bibitem[{{Gorbovskoy} {et~al}\mbox{.}(2010){Gorbovskoy}, {Lipunov},
  {Kornilov}, {Belinski}, {Shatskiy}, {Tyurina}, {Kuvshinov}, {Balanutsa},
  {Chazov}, {Kortunov}, {Kuznetsov}, \& {et al}}]{gcn11185}
{Gorbovskoy} E. {et~al.}, 2010, GRB Coordinates Network, Circular Service, 1185

\bibitem[{{Grupe} {et~al}\mbox{.}(2007){Grupe}, {Nousek}, {vanden Berk},
  {Roming}, {Burrows}, {Godet}, {Osborne}, \& {Gehrels}}]{grupe_etal2007}
{Grupe} D., {Nousek} J.~A., {vanden Berk} D.~E., {Roming} P.~W.~A., {Burrows}
  D.~N., {Godet} O., {Osborne} J., {Gehrels} N., 2007, \aj, 133, 2216

\bibitem[{{Gruzinov} \& {Waxman}(1999)}]{gruzinov-waxman1999}
{Gruzinov} A., {Waxman} E., 1999, \apj, 511, 852

\bibitem[{{Hayes}(1985)}]{Hayes1985}
{Hayes} D.~S., 1985, in IAU Symposium, Vol. 111, Calibration of Fundamental
  Stellar Quantities, {D.~S.~Hayes, L.~E.~Pasinetti, \& A.~G.~D.~Philip}, ed.,
  pp. 225--249

\bibitem[{{Horv{\'a}th}(1998)}]{horvath1998}
{Horv{\'a}th} I., 1998, \apj, 508, 757

\bibitem[{{Horv{\'a}th}(2002)}]{horvath2002}
---, 2002, \aap, 392, 791

\bibitem[{{Immler} {et~al}\mbox{.}(2010){Immler}, {Barthelmy}, {Baumgartner},
  {Beardmore}, {Campana}, {D'Elia}, {Evans}, {Gelbord}, {Godet}, {Gronwall},
  {Guidorzi}, \& {et al}}]{gcn11159}
{Immler} S. {et~al.}, 2010, GRB Coordinates Network, Circular Service, 11159, 1
  (2010), 1159, 1

\bibitem[{{Ivanov} {et~al}\mbox{.}(2010{\natexlab{a}}){Ivanov}, {Chuvalaev},
  {Poleschuk}, {Konstantinov}, {Lenok}, {Gres}, {Yazev}, {Budnev},
  {Gorbovskoy}, {Lipunov}, {Kornilov}, {Belinski}, \& {et al}}]{gcn11161}
{Ivanov} K. {et~al.}, 2010{\natexlab{a}}, GRB Coordinates Network, Circular
  Service, 11161, 1 (2010), 1161, 1

\bibitem[{{Ivanov} {et~al}\mbox{.}(2010{\natexlab{b}}){Ivanov}, {Chuvalaev},
  {Poleschuk}, {Konstantinov}, {Lenok}, {Gres}, {Yazev}, {Budnev},
  {Gorbovskoy}, {Lipunov}, {Kornilov}, {Belinski}, \& {et al}}]{gcn11163}
---, 2010{\natexlab{b}}, GRB Coordinates Network, Circular Service, 11163, 1
  (2010), 1163, 1

\bibitem[{{Ivanov} {et~al}\mbox{.}(2010{\natexlab{c}}){Ivanov}, {Chuvalaev},
  {Poleschuk}, {Konstantinov}, {Lenok}, {Gres}, {Yazev}, {Budnev},
  {Gorbovskoy}, {Lipunov}, {Kornilov}, {Belinski}, \& {et al}}]{gcn11216}
---, 2010{\natexlab{c}}, GRB Coordinates Network, Circular Service, 11216, 1
  (2010), 1216, 1

\bibitem[{{Jakobsson} {et~al}\mbox{.}(2004){Jakobsson}, {Hjorth}, {Fynbo},
  {Watson}, {Pedersen}, {Bj{\"o}rnsson}, \& {Gorosabel}}]{Jakobsson_etal2004}
{Jakobsson} P., {Hjorth} J., {Fynbo} J.~P.~U., {Watson} D., {Pedersen} K.,
  {Bj{\"o}rnsson} G., {Gorosabel} J., 2004, \apjl, 617, L21

\bibitem[{{Jensen} {et~al}\mbox{.}(2001){Jensen}, {Fynbo}, {Gorosabel},
  {Hjorth}, {Holland}, {M{\"o}ller}, {Thomsen}, {Bj{\"o}rnsson}, {Pedersen},
  {Burud}, {Henden}, {Tanvir}, {Davis}, {Vreeswijk}, {Rol}, {Hurley}, {Cline},
  {Trombka}, {McClanahan}, {Starr}, {Goldsten}, {Castro-Tirado}, {Greiner},
  {Bailer-Jones}, {K{\"u}mmel}, \& {Mundt}}]{Jensen_etal2001}
{Jensen} B.~L. {et~al.}, 2001, \aap, 370, 909

\bibitem[{{Kalberla} {et~al}\mbox{.}(2005){Kalberla}, {Burton}, {Hartmann},
  {Arnal}, {Bajaja}, {Morras}, \& {P{\"o}ppel}}]{kalberla_etal2005}
{Kalberla} P.~M.~W., {Burton} W.~B., {Hartmann} D., {Arnal} E.~M., {Bajaja} E.,
  {Morras} R., {P{\"o}ppel} W.~G.~L., 2005, \aap, 440, 775

\bibitem[{{Katz} \& {Piran}(1997)}]{katz-piran1997}
{Katz} J.~I., {Piran} T., 1997, \apj, 490, 772

\bibitem[{{Kornilov} {et~al}\mbox{.}(2011){Kornilov}, {Lipunov}, {Gorbovskoy},
  {Belinski}, {Kuvshinov}, {Tyurina}, {Sankovich}, {Krylov}, {Shatsky},
  {Balanutsa}, {Chazov}, {Kuznetsov}, {Zimnuhov}, {Senik}, {Tlatov},
  {Parkhomenko}, {Dormidontov}, {Krushinsky}, {Zalozhnyh}, {Popov}, {Yazev},
  {Budnev}, {Ivanov}, {Konstantinov}, {Gress}, {Chvalaev}, {Yurkov},
  {Sergienko}, \& {Kudelina}}]{kornilov_etal2011}
{Kornilov} V.~G. {et~al.}, 2011, Experimental Astronomy, accepted for
  publication

\bibitem[{{Krushinski} {et~al}\mbox{.}(2010{\natexlab{a}}){Krushinski},
  {Zalozhnich}, {Kopytova}, {Popov}, {Gorbovskoy}, {Lipunov}, {Kornilov},
  {Belinski}, {Shatskiy}, {Tyurina}, {Kuvshinov}, \& {et al}}]{gcn11182}
{Krushinski} V. {et~al.}, 2010{\natexlab{a}}, GRB Coordinates Network, Circular
  Service, 11182, 1 (2010), 1182, 1

\bibitem[{{Krushinski} {et~al}\mbox{.}(2010{\natexlab{b}}){Krushinski},
  {Zalozhnich}, {Kopytova}, {Popov}, {Ivanov}, {Chuvalaev}, {Poleschuk},
  {Konstantinov}, {Lenok}, {Gres}, {Yazev}, {Budnev}, {Gorbovskoy}, {Lipunov},
  {Kornilov}, \& {et al}}]{gcn11359}
---, 2010{\natexlab{b}}, GRB Coordinates Network, Circular Service, 11359, 1
  (2010), 1359, 1

\bibitem[{{Krushinski} {et~al}\mbox{.}(2010{\natexlab{c}}){Krushinski},
  {Zalozhnich}, {Kopytova}, {Popov}, {Ivanov}, {Chuvalaev}, {Poleschuk},
  {Konstantinov}, {Lenok}, {Gres}, {Yazev}, {Budnev}, {Gorbovskoy}, {Lipunov},
  {Kornilov}, \& {et al}}]{gcn11361}
---, 2010{\natexlab{c}}, GRB Coordinates Network, Circular Service, 11361, 1
  (2010), 1361, 1

\bibitem[{{Kulkarni} {et~al}\mbox{.}(1998){Kulkarni}, {Frail}, {Wieringa},
  {Ekers}, {Sadler}, {Wark}, {Higdon}, {Phinney}, \&
  {Bloom}}]{Kulkarni_etal1998}
{Kulkarni} S.~R. {et~al.}, 1998, \nat, 395, 663

\bibitem[{{Kuvshinov} {et~al}\mbox{.}(2010){Kuvshinov}, {Kornilov},
  {Gorbovskoy}, {Lipunov}, {Belinski}, {Shatskiy}, {Tyurina}, {Balanutsa},
  {Chazov}, {Kortunov}, \& et~al}]{gcn11235}
{Kuvshinov} D. {et~al.}, 2010, GRB Coordinates Network, Circular Service,
  11235, 1 (2010), 1235, 1

\bibitem[{{Landi Degl'Innocenti}, {Bagnulo} \& {Fossati}(2007){Landi
  Degl'Innocenti}, {Bagnulo}, \& {Fossati}}]{landi_etal2007}
{Landi Degl'Innocenti} E., {Bagnulo} S., {Fossati} L., 2007, in Astronomical
  Society of the Pacific Conference Series, Vol. 364, The Future of
  Photometric, Spectrophotometric and Polarimetric Standardization,
  {C.~Sterken}, ed., pp. 495--502

\bibitem[{{Liang}, {Zhang} \& {Zhang}(2007){Liang}, {Zhang}, \&
  {Zhang}}]{liang_etal2007}
{Liang} E.-W., {Zhang} B.-B., {Zhang} B., 2007, \apj, 670, 565

\bibitem[{{Lipunov} \& {Gorbovskoy}(2007)}]{lipunov-gorbovskoy2007}
{Lipunov} V., {Gorbovskoy} E., 2007, \apjl, 665, L97

\bibitem[{{Lipunov} {et~al}\mbox{.}(2010){Lipunov}, {Kornilov}, {Gorbovskoy},
  {Shatskij}, {Kuvshinov}, {Tyurina}, {Belinski}, {Krylov}, \& {et
  al}}]{lipunov_etal2010}
{Lipunov} V. {et~al.}, 2010, Advances in Astronomy, article id. 349171

\bibitem[{{Lipunov} \& {Gorbovskoy}(2008)}]{lipunov-gorb2008}
{Lipunov} V.~M., {Gorbovskoy} E.~S., 2008, \mnras, 383, 1397

\bibitem[{{Lipunova}(1997)}]{lipunova1997e}
{Lipunova} G.~V., 1997, Astronomy Letters, 23, 84

\bibitem[{{Markwardt} {et~al}\mbox{.}(2010){Markwardt}, {Barthelmy},
  {Beardmore}, {Burrows}, {D'Elia}, {Evans}, {Gehrels}, {Gelbord}, {Godet},
  {Kennea}, {Kuin}, \& {et al}}]{gcn11227}
{Markwardt} C.~B. {et~al.}, 2010, GRB Coordinates Network, Circular Service,
  11227, 1 (2010), 1227, 1

\bibitem[{{Marshall} {et~al}\mbox{.}(2010){Marshall}, {Barthelmy}, {Beardmore},
  {Burrows}, {Evans}, {Gehrels}, {Gelbord}, {Guidorzi}, {Holland}, {O'Brien},
  {Page}, {Palmer}, {Rowlinson}, {Sbarufatti}, {Siegel}, {Troja}, \&
  {Ukwatta}}]{gcn11214}
{Marshall} F.~E. {et~al.}, 2010, GRB Coordinates Network, Circular Service,
  11214, 1 (2010), 1214, 1

\bibitem[{{M{\'e}sz{\'a}ros}(2002)}]{meszaros2002}
{M{\'e}sz{\'a}ros} P., 2002, \araa, 40, 137

\bibitem[{{Meszaros} \& {Rees}(1997)}]{meszaros-rees1997}
{Meszaros} P., {Rees} M.~J., 1997, \apj, 476, 232

\bibitem[{{M{\'e}sz{\'a}ros} \& {Rees}(1999)}]{meszaros-rees1999}
{M{\'e}sz{\'a}ros} P., {Rees} M.~J., 1999, \mnras, 306, L39

\bibitem[{{Metzger}(2010)}]{Metzger2010}
{Metzger} B.~D., 2010, in Astronomical Society of the Pacific Conference
  Series, Vol. 432, New Horizons in Astronomy: Frank N. Bash Symposium 2009,
  {L.~M.~Stanford, J.~D.~Green, L.~Hao, \& Y.~Mao}, ed., pp. 82--96

\bibitem[{{Mukherjee} {et~al}\mbox{.}(1998){Mukherjee}, {Feigelson}, {Jogesh
  Babu}, {Murtagh}, {Fraley}, \& {Raftery}}]{Mukherjee_etal1998}
{Mukherjee} S., {Feigelson} E.~D., {Jogesh Babu} G., {Murtagh} F., {Fraley} C.,
  {Raftery} A., 1998, \apj, 508, 314

\bibitem[{{Mundell} {et~al}\mbox{.}(2007){Mundell}, {Steele}, {Smith},
  {Kobayashi}, {Melandri}, {Guidorzi}, {Gomboc}, {Mottram}, {Clarke},
  {Monfardini}, {Carter}, \& {Bersier}}]{mundell_etal2007}
{Mundell} C.~G. {et~al.}, 2007, Science, 315, 1822

\bibitem[{{Page} \& {Immler}(2010)}]{gcn11171}
{Page} K.~L., {Immler} S., 2010, GRB Coordinates Network, Circular Service,
  11171, 1 (2010), 1171, 1

\bibitem[{{Pei}(1992)}]{Pei1992}
{Pei} Y.~C., 1992, \apj, 395, 130

\bibitem[{{Pickles} \& {Depagne}(2010)}]{Pickles-Depagne2010}
{Pickles} A., {Depagne} {\'E}., 2010, \pasp, 122, 1437

\bibitem[{{Piran}(2005)}]{piran2005}
{Piran} T., 2005, Reviews of Modern Physics, 76, 1143

\bibitem[{{Racusin} {et~al}\mbox{.}(2009){Racusin}, {Liang}, {Burrows},
  {Falcone}, {Sakamoto}, {Zhang}, {Zhang}, {Evans}, \&
  {Osborne}}]{racusin_etal2009}
{Racusin} J.~L. {et~al.}, 2009, \apj, 698, 43

\bibitem[{{Sakamoto} {et~al}\mbox{.}(2010{\natexlab{a}}){Sakamoto},
  {Barthelmy}, {Baumgartner}, {Cummings}, {Fenimore}, {Gehrels}, {Krimm},
  {Markwardt}, {Palmer}, {Sato}, {Saxton}, {Stamatikos}, {Tueller}, \&
  {Ukwatta}}]{gcn11358a}
{Sakamoto} T. {et~al.}, 2010{\natexlab{a}}, GRB Coordinates Network, Circular
  Service, 11358, 1 (2010), 1358, 1

\bibitem[{{Sakamoto} {et~al}\mbox{.}(2010{\natexlab{b}}){Sakamoto},
  {Barthelmy}, {Baumgartner}, {Cummings}, {Gehrels}, {Krimm}, {Immler},
  {Markwardt}, {Palmer}, {Stamatikos}, {Tueller}, \& {Ukwatta}}]{gcn11169b}
---, 2010{\natexlab{b}}, GRB Coordinates Network, Circular Service, 11169, 1
  (2010), 1169, 1

\bibitem[{{Sakamoto} {et~al}\mbox{.}(2010{\natexlab{c}}){Sakamoto},
  {Baumgartner}, {Beardmore}, {Cummings}, {Evans}, {Gehrels}, {Gelbord}, \& {et
  al}}]{gcn11181c}
{Sakamoto} T., {Baumgartner} W.~H., {Beardmore} A.~P., {Cummings} J.~R.,
  {Evans} P.~A., {Gehrels} N., {Gelbord} J.~M., {et al}, 2010{\natexlab{c}},
  GRB Coordinates Network, Circular Service, 11181, 1 (2010), 1181, 1

\bibitem[{{Sanchez-Ramirez} {et~al}\mbox{.}(2010){Sanchez-Ramirez}, {Tello},
  {Sota}, {Gorosabel}, \& {Castro-Tirado}}]{gcn11180}
{Sanchez-Ramirez} R., {Tello} J.~C., {Sota} A., {Gorosabel} J., {Castro-Tirado}
  A.~J., 2010, GRB Coordinates Network, Circular Service, 11180, 1 (2010),
  1180, 1

\bibitem[{{Sari} \& {Piran}(1997)}]{sari-piran1997}
{Sari} R., {Piran} T., 1997, \apj, 485, 270

\bibitem[{{Sari} \& {Piran}(1999)}]{sari-piran1999}
---, 1999, \apj, 520, 641

\bibitem[{{Sari}, {Piran} \& {Narayan}(1998){Sari}, {Piran}, \&
  {Narayan}}]{sari_etal1998}
{Sari} R., {Piran} T., {Narayan} R., 1998, \apjl, 497, L17

\bibitem[{{Saxton} {et~al}\mbox{.}(2010){Saxton}, {Baumgartner}, {Beardmore},
  {Cummings}, {de Pasquale}, {Gehrels}, {Gelbord}, {Holland}, {Kennea},
  {Krimm}, {Kuin}, {Littlejohns}, {Markwardt}, {Page}, {Palmer}, {Siegel}, \&
  {Troja}}]{gcn11357}
{Saxton} C.~J. {et~al.}, 2010, GRB Coordinates Network, Circular Service,
  11357, 1 (2010), 1357, 1

\bibitem[{{Schady} {et~al}\mbox{.}(2010){Schady}, {Page}, {Oates}, {Still}, {de
  Pasquale}, {Dwelly}, {Kuin}, {Holland}, {Marshall}, \&
  {Roming}}]{Schady_etal2010}
{Schady} P. {et~al.}, 2010, \mnras, 401, 2773

\bibitem[{{Schlegel}, {Finkbeiner} \& {Davis}(1998){Schlegel}, {Finkbeiner}, \&
  {Davis}}]{Schlegel_etal1998}
{Schlegel} D.~J., {Finkbeiner} D.~P., {Davis} M., 1998, \apj, 500, 525

\bibitem[{{Siegel} \& {Marshall}(2010)}]{gcn11237}
{Siegel} M.~H., {Marshall} F.~E., 2010, GRB Coordinates Network, Circular
  Service, 11237, 1 (2010), 1237, 1

\bibitem[{{Sota}, {de Ugarte Postigo} \& {Castro-Tirado}(2010){Sota}, {de
  Ugarte Postigo}, \& {Castro-Tirado}}]{gcn11220}
{Sota} A., {de Ugarte Postigo} A., {Castro-Tirado} A.~J., 2010, GRB Coordinates
  Network, Circular Service, 11220, 1 (2010), 1220, 1

\bibitem[{{Stamatikos} {et~al}\mbox{.}(2010){Stamatikos}, {Barthelmy},
  {Baumgartner}, {Cummings}, {Gehrels}, {Krimm}, {Markwardt}, {Palmer},
  {Sakamoto}, {Tueller}, \& {Ukwatta}}]{gcn11202}
{Stamatikos} M. {et~al.}, 2010, GRB Coordinates Network, Circular Service,
  11202, 1 (2010), 1202, 1

\bibitem[{{Steele} {et~al}\mbox{.}(2009){Steele}, {Mundell}, {Smith},
  {Kobayashi}, \& {Guidorzi}}]{steele_etal2009}
{Steele} I.~A., {Mundell} C.~G., {Smith} R.~J., {Kobayashi} S., {Guidorzi} C.,
  2009, \nat, 462, 767

\bibitem[{{Tanvir}, {Wiersema} \& {Levan}(2010){Tanvir}, {Wiersema}, \&
  {Levan}}]{gcn11230}
{Tanvir} N.~R., {Wiersema} K., {Levan} A.~J., 2010, GRB Coordinates Network,
  Circular Service, 11230, 1 (2010), 1230, 1

\bibitem[{{Tello} {et~al}\mbox{.}(2010){Tello}, {Sanchez-Ramirez}, {Sota},
  {Gorosabel}, \& {Castro-Tirado}}]{gcn11196}
{Tello} J.~C., {Sanchez-Ramirez} R., {Sota} A., {Gorosabel} J., {Castro-Tirado}
  A.~J., 2010, GRB Coordinates Network, Circular Service, 11196, 1 (2010),
  1196, 1

\bibitem[{{Tody}(1993)}]{tody1993}
{Tody} D., 1993, in Astronomical Society of the Pacific Conference Series,
  Vol.~52, Astronomical Data Analysis Software and Systems II, {R.~J.~Hanisch,
  R.~J.~V.~Brissenden, \& J.~Barnes}, ed., pp. 173--183

\bibitem[{{Troja} {et~al}\mbox{.}(2007){Troja}, {Cusumano}, {O'Brien}, {Zhang},
  {Sbarufatti}, {Mangano}, {Willingale}, {Chincarini}, {Osborne}, {Marshall},
  {Burrows}, {Campana}, {Gehrels}, {Guidorzi}, {Krimm}, {La Parola}, {Liang},
  {Mineo}, {Moretti}, {Page}, {Romano}, {Tagliaferri}, {Zhang}, {Page}, \&
  {Schady}}]{troja_etal2007}
{Troja} E. {et~al.}, 2007, \apj, 665, 599

\bibitem[{{Vestrand} {et~al}\mbox{.}(2005){Vestrand}, {Wozniak}, {Wren},
  {Fenimore}, {Sakamoto}, {White}, {Casperson}, {Davis}, {Evans}, {Galassi}, \&
  {et al.}}]{vestrand_etal2005}
{Vestrand} W.~T. {et~al.}, 2005, \nat, 435, 178

\bibitem[{{Vestrand} {et~al}\mbox{.}(2006){Vestrand}, {Wren}, {Wozniak},
  {Aptekar}, {Golentskii}, {Pal'Shin}, {Sakamoto}, {White}, {Evans},
  {Casperson}, \& {Fenimore}}]{vestrand_etal2006}
---, 2006, \nat, 442, 172

\bibitem[{{Willingale} {et~al}\mbox{.}(2007){Willingale}, {O'Brien}, {Osborne},
  {Godet}, {Page}, {Goad}, {Burrows}, {Zhang}, {Rol}, {Gehrels}, \&
  {Chincarini}}]{willingale_etal2007}
{Willingale} R. {et~al.}, 2007, \apj, 662, 1093

\bibitem[{{Winkler} {et~al}\mbox{.}(2003){Winkler}, {Courvoisier}, {Di Cocco},
  {Gehrels}, {Gim{\'e}nez}, {Grebenev}, {Hermsen}, {Mas-Hesse}, {Lebrun},
  {Lund}, {Palumbo}, {Paul}, {Roques}, {Schnopper}, {Sch{\"o}nfelder},
  {Sunyaev}, {Teegarden}, {Ubertini}, {Vedrenne}, \& {Dean}}]{Integral2003}
{Winkler} C. {et~al.}, 2003, \aap, 411, L1

\bibitem[{{Zhang} {et~al}\mbox{.}(2006){Zhang}, {Fan}, {Dyks}, {Kobayashi},
  {M{\'e}sz{\'a}ros}, {Burrows}, {Nousek}, \& {Gehrels}}]{zhang_etal2006}
{Zhang} B., {Fan} Y.~Z., {Dyks} J., {Kobayashi} S., {M{\'e}sz{\'a}ros} P.,
  {Burrows} D.~N., {Nousek} J.~A., {Gehrels} N., 2006, \apj, 642, 354

\bibitem[{{Zhang} \& {M{\'e}sz{\'a}ros}(2001)}]{zhang-meszaros2001}
{Zhang} B., {M{\'e}sz{\'a}ros} P., 2001, \apjl, 552, L35

\bibitem[{{Zhang} \& {M{\'e}sz{\'a}ros}(2004)}]{zhang-meszaros2004}
---, 2004, International Journal of Modern Physics A, 19, 2385

\bibitem[{{Zhang}(2009)}]{Zhang2009}
{Zhang} X.-H., 2009, Research in Astronomy and Astrophysics, 9, 213

\end{thebibliography}

\begin{figure}
\includegraphics[width=80mm]{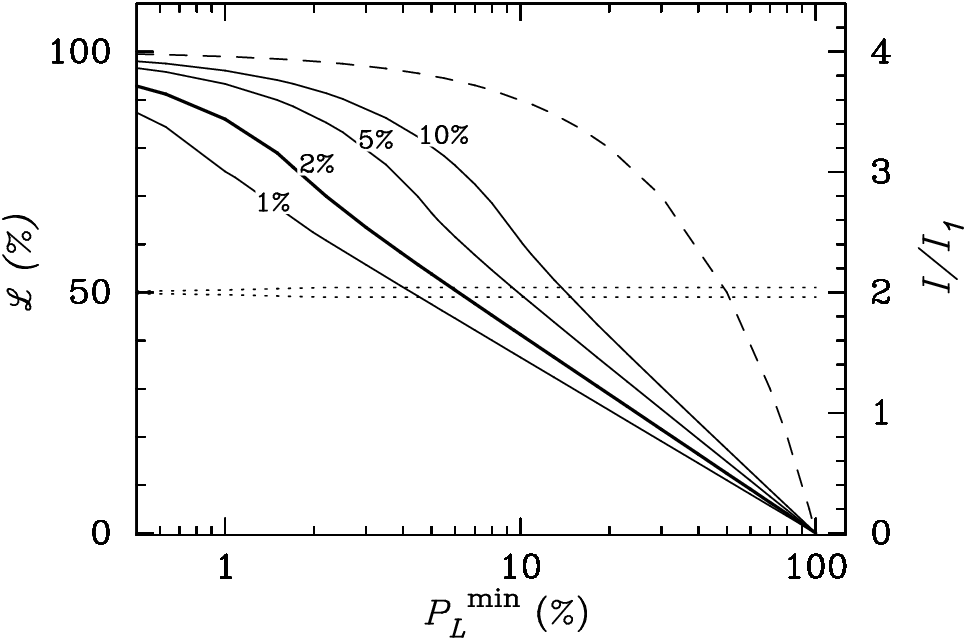} \caption{ Left axis: 
likelihood $\mathcal{L}(P_L\geq P_L^{\mathrm{min}}\mid I_1\approx I_2)$,
where $I_{1,2}$ are the signals from two orthogonal polarizing filters,
$P_L$ is an unknown fractional linear polarization of a source, and
$P_L^{\mathrm{min}}$ is a value plotted along the $x$-axis. The several
likelihood functions are plotted as solid curves marked with the values
of $\sigma_\mathrm{p}$, the known relative error between the
measurements with two orthogonal polarizing filters. The dashed line
shows the probability of an unobserved source to have $P_\mathrm{L}\geq
P_L^\mathrm{min}$ under the prior hypothesis that a source has no
preference to any degree of linear polarization. Right axis: the ratio
of the total intensity $I$ to the measured signal $I_1$ for
$|D|<\sigma_\mathrm{p}=2\%$. Its limits are shown by the double dotted
line.
\label{fig:prob_polarization} }
 \end{figure}

\appendix 

\subsection*{Appendix. Estimating degree of linear polarization from observations
with two fixed polarizing filters\label{section:polarization}}
Observations of bright GRB\,100906 have been made with just one 
telescope able to provide polarization measurements (the MASTER
telescope at Tunka). Generally, if an observation is made only with two
polarizing filters at a single set of positions, one can obtain just an
estimate on the degree of linear polarization.

In the consideration below we assume zero circular polarization. Let
$P_L$ be the fraction of linearly polarized radiation and $\theta$ is
the angle of maximum polarization, i.e. the angle between the $x$-axis
of the reference system and a segment which a polarization ellipse
degenerates into for the case with zero circular polarization (see,
e.g., \cite{landi_etal2007}). Then $P_L=\sqrt{P_Q^2+P_U^2}$, where
$P_Q=Q/I$, $P_U=U/I$, and $I$, $Q$, $U$ are the first three Stokes
parameters; $I$ is the total intensity with no polarization filter
interposed (in this section we do not consider color filters).

To derive value of $P_L$ one needs to perform observations with filters
for linear polarization positioned at three angles, for example, at
$0^o$, $45^o$, and $90^o$ with respect to the reference direction. We
have  two orthogonal filters at each of the MASTER-II telescopes; thus, if
such telescope observes alone, only a lower limit on $P_L$ can be
deduced. If this lower limit is a virtual zero, there is still some
probability that actual linear polarization is substantial. 

Let $I_1$ and $I_2$ be the  signals of the detector with filters at
$0^o$ and $90^o$ interposed. Then a value $D=(I_1-I_2)/(I_1+I_2)$ can
be calculated, which is expressed as follows:
\begin{equation}
D = P_L \, \cos 2\theta \, .
\label{eq:D_polariz}
\end{equation} 

If the value of $D$ differs from zero ($|D|$ is greater than the
relative error $\sigma_\mathrm{p}$ between measurements with two
differently positioned filters), a certain amount of linear 
polarization is present in the light with a lower estimate for it:
$P_L^\mathrm{min} = |D|$, calculated as if a filter is interposed at
angle $0^o$ or $90^o$ with respect to the direction of polarization
plane of the incident light.

If the module of $D$ is less than $\sigma_\mathrm{p}$, then formally we
have zero as a minimum estimate for the degree of linear polarization.
100-percent linear polarization is also possible in this case, if a
filter is accidentally positioned so that a source's polarization plane
is inclined by $45^o$ to it. This possibility can be numerically
evaluated considering a total range of the filter positions that would
result in a value of $|D|<\sigma_\mathrm{p}$. Substituting
$D=\sigma_\mathrm{p}$ and $\theta = \pi/4+\delta\theta/2$ in
eq.~(\ref{eq:D_polariz}), for any preset value of $P_\mathrm{L}$ we get
a value of the allowable interval $\delta\theta$. As our configuration
consists of two orthogonal filters, source polarization plane should be
in either of four intervals $\delta \theta$ around angles $N \times
\pi/4$, where $N=1,2,3,4$. Thus, a source with linear polarization
degree $P_\mathrm{L}$ yields value $|D|<\sigma_\mathrm{p}$ with
probability $4\, \delta \theta/2\pi$. Now we can calculate the
conditional probability of a source with observed
$|D|<\sigma_\mathrm{p}$ to have a particular degree of linear
polarization $P_\mathrm{L}$ using the Bayes' theorem. We make a
conservative assumption that a source can have any fractional linear
polarization from 0 to 100\%. From this, the likelihood function
$\mathcal{L}(P_L>P_L^{\mathrm{min}}\mid I_1\approx
I_2)=\mathcal{L}(P_L>P_L^{\mathrm{min}}\mid \, |D|<\sigma_\mathrm{p})$
can be derived, where $I_1\approx I_2$ is a realized event, $P_L$ is the
actual fractional linear polarization of a source, and
$P_L^\mathrm{min}$ is some value of it. The probability that the actual
linear polarization of a source exceeds value $P_L$ is shown in 
Fig.~\ref{fig:prob_polarization}. Thus, for example, if no difference
between $I_1$ and $I_2$ is registered in the measurements with 2 per
cent error, there is a fifty-percent chance that a source has linear
degree of polarization greater than $\approx 6$ per cent.

Accordingly, a total intensity from the source is greater than observed
$I_1\approx I_2$. One can see from Fig.~\ref{fig:prob_polarization} that
in the observations with $|D|<\sigma_\mathrm{p}$ the total intensity is
always about the same factor greater (twice) than measured $I_1$. This
can be seen considering expression for the signal from one polarizer:
$I_1 = I\,(1-P_L)/2+ I\,P_L\cos^2\theta$. For small $P_L$ the light behaves
as unpolarized, and for large $P_L$ the allowable interval
$\delta\theta$ approaches zero and $\theta \approx \pi/4$.

\end{document}